\documentclass[sigconf]{acmart}

\setcopyright{none}
\renewcommand\footnotetextcopyrightpermission[1]{}

\newif\ifarxiv
\arxivtrue

\ifarxiv
\renewcommand\footnotetextcopyrightpermission[1]{} 
\fi

\def\BibTeX{{\rm B\kern-.05em{\sc i\kern-.025em b}\kern-.08emT\kern-.1667em\lower.7ex\hbox{E}\kern-.125emX}}
 
\usepackage{graphicx}
\usepackage{balance}

\usepackage{etex}

\usepackage{colortbl}

\usepackage[font={sf,small}]{caption}
\usepackage[font={sf,small}]{subcaption}

\usepackage[sc]{mathpazo}

\usepackage{graphicx}
\usepackage{booktabs}
\usepackage{epstopdf}
\usepackage{url}
\usepackage{placeins}
\usepackage{multirow}
\usepackage{rotating}
\usepackage{pbox}
\usepackage[normalem]{ulem}
\usepackage{listings}

\usepackage{cleveref}
\usepackage[utf8]{inputenc}
\crefname{section}{§}{§§}
\Crefname{section}{§}{§§}

\usepackage[10pt]{moresize}

\usepackage{xcolor}
\definecolor{darkgrey}{RGB}{70,70,70}
\definecolor{lightgrey}{RGB}{200,200,200}
\definecolor{lyellow}{RGB}{255,255,100}
\definecolor{llyellow}{RGB}{250,250,180}
\definecolor{lgreen}{RGB}{144,238,144}

\usepackage{enumitem}

\usepackage{listings}

\usepackage[linesnumbered,ruled,vlined]{algorithm2e}
\SetAlFnt{\scriptsize\tt}
\SetAlCapFnt{\scriptsize\sf}
\SetAlCapNameFnt{\scriptsize\sf}

\newcommand{\maciej}[1]{\textcolor{blue}{[Maciej: #1]}}

\newcommand{\lukas}[1]{\textcolor{blue}{[Lukas: #1]}}
\newcommand{\andrei}[1]{\textcolor{blue}{[Andrei: #1]}}

\newcommand{\macb}[1]{\textbf{\textsf{#1}}}

\usepackage{scalerel,stackengine}
\stackMath
\newcommand\rwh[1]{%
\savestack{\tmpbox}{\stretchto{%
  \scaleto{%
        \scalerel*[\widthof{\ensuremath{#1}}]{\kern-.6pt\bigwedge\kern-.6pt}%
                  {\rule[-\textheight/2]{1ex}{\textheight}}
                              }{\textheight}%
}{0.5ex}}%
\stackon[1pt]{#1}{\tmpbox}%
}

\usepackage[hang,flushmargin]{footmisc}

\usepackage{soul}

\usepackage{fontawesome}
\usepackage{pifont}
\usepackage{textcomp}

\usepackage{makecell}

\usepackage{lscape}

\usepackage{tikz}
\usetikzlibrary{tikzmark}

\usepackage{xpatch}
\expandafter\xpatchcmd
\csname pgfk@/tikz/every picture/.@cmd\endcsname
{\thepage}{\arabic{page}}{}{}

\tikzstyle{comment} = [draw, fill=blue!70, text=white, text width=3cm, minimum height=1cm, rounded corners, align=left, font=\scriptsize]
\tikzstyle{background_alg} = [draw, fill=blue!20, opacity=0.4, inner sep=4pt, rounded corners=2pt]

\usetikzlibrary{shapes}
\usetikzlibrary{plotmarks}
\usetikzlibrary{calc,fit}

\DeclareRobustCommand{\hll}[1]{{\sethlcolor{llyellow}\hl{#1}}}

\newcommand{\add}{\hspace{19em}\makebox[0pt]{\color{llyellow}\rule[-0.1ex]{39em}{2ex}}\hspace{-19em}} 

\newif\iftr
\newif\ifsc
\newif\ifscr

\newif\ifconf

\conffalse

\sctrue
\scrtrue

\ifconf

\setcopyright{acmlicensed}

\fi

\ifarxiv

\setcopyright{none}

\fi

\renewcommand{\textrightarrow}{$\rightarrow$}

\begin{document}

\ifconf

\copyrightyear{2019} 
\acmYear{2019} 
\acmConference[SC '19]{The International Conference for High Performance Computing, Networking, Storage, and Analysis}{November 17--22, 2019}{Denver, CO, USA}
\acmBooktitle{The International Conference for High Performance Computing, Networking, Storage, and Analysis (SC '19), November 17--22, 2019, Denver, CO, USA}
\acmPrice{15.00}
\acmDOI{10.1145/3295500.3356182}
\acmISBN{978-1-4503-6229-0/19/11}

\fi

\title[Slim Graph: Practical Lossy Graph Compression for Approximate Graph Computations]{Slim Graph: Practical Lossy Graph Compression\\ for Approximate Graph Processing, Storage, and Analytics}

\author{Maciej Besta, Simon Weber, Lukas Gianinazzi, Robert Gerstenberger,\break Andrey Ivanov, Yishai Oltchik, Torsten Hoefler}
\affiliation{\vspace{0.25em}Department of Computer Science; ETH Zurich}

%
%

\renewcommand{\shortauthors}{M. Besta, S. Weber, L. Gianinazzi, R. Gerstenberger, A. Ivanov, Y. Oltchik, T. Hoefler}

\begin{abstract}
We propose Slim Graph: the first programming model and framework for practical
lossy graph compression that facilitates high-performance approximate graph
processing, storage, and analytics. Slim Graph enables the developer to express
numerous compression schemes using small and programmable compression kernels
that can access and modify local parts of input graphs. Such kernels are
executed in parallel by the underlying engine, isolating developers
from complexities of parallel programming. Our kernels implement novel graph
compression schemes that preserve numerous graph properties, for example
connected components, minimum spanning trees, or graph spectra.  Finally, Slim
Graph uses statistical divergences and other metrics to analyze the accuracy of
lossy graph compression. We illustrate both theoretically and empirically that
Slim Graph accelerates numerous graph algorithms, reduces storage used by graph
datasets, and ensures high accuracy of results. Slim Graph may become the
common ground for developing, executing, and analyzing emerging lossy graph
compression schemes.

\end{abstract}



%


\maketitle

\thispagestyle{empty}

\ifsc
{\small\noindent\macb{Slim Graph website:}\\\url{https://github.com/mbesta/SlimGraph}}
\fi


\section{Introduction}
\label{sec:intro}


Large graphs are a basis of many problems in machine learning, medicine, social
network analysis, computational sciences, and
others~\cite{DBLP:journals/ppl/LumsdaineGHB07, ben2019modular,
besta2019communication}.  The growing graph sizes, reaching one trillion edges
in 2015 (the Facebook social graph~\cite{ching2015one}) and 12 trillion edges
in 2018 (the Sogou webgraph~\cite{lin2018shentu}), require unprecedented
amounts of compute power, storage, and energy. For example, running PageRank on
the Sogou webgraph using 38,656 compute nodes (10,050,560 cores) on the Sunway
TaihuLight supercomputer~\cite{fu2016sunway} (nearly the full scale of
TaihuLight) takes 8 minutes~\cite{lin2018shentu}.  The sizes of such datasets
will continue to grow; Sogou Corp.~expects a $\approx$60 trillion edge graph
dataset with whole-web crawling.
{Lowering the size of such graphs is increasingly important for academia
and industry}: It would offer speedups by reducing the number of expensive I/O
operations, the amount of data communicated over the
network~\cite{besta2014slim, besta2018slim, besta2019fatpaths} and by storing a
larger fraction of data in caches.

%
There exist many \emph{lossless} schemes for graph compression, including
WebGraph~\cite{boldi2004webgraph}, $k^2$-trees~\cite{brisaboa2009k2}, and
others~\cite{besta2018survey}.
%
%
They provide various degrees of storage reductions. Unfortunately, the majority
of these schemes incur \emph{expensive decompression} in performance-critical
kernels and \emph{high preprocessing costs} that throttle
performance~\cite{boldi2004webgraph, brisaboa2009k2}.
Moreover, there also exist \emph{succinct} graph representations that
{approach the associated graph storage lower
bounds}~\cite{turan1984succinct, naor1990succinct, raman2007succinct,
farzan2008succinct}. However, they are mostly theoretical structures with large
hidden constants. In addition, as shown recently, the associated storage
reductions are not large, at most 20--35\%, as {today's graph codes
already come close to theoretical storage lower bounds}~\cite{besta2018log}.

\begin{figure*}[hbtp]
  \vspace{4.5em}
  \centering
\textbf{Lossy compression of bitmaps (JPEG):}\\
  \begin{subfigure}[t]{0.4 \textwidth}
  \captionsetup{labelfont={color=white, scriptsize},font={color=white, scriptsize}}
    \centering
    \includegraphics[width=\textwidth]{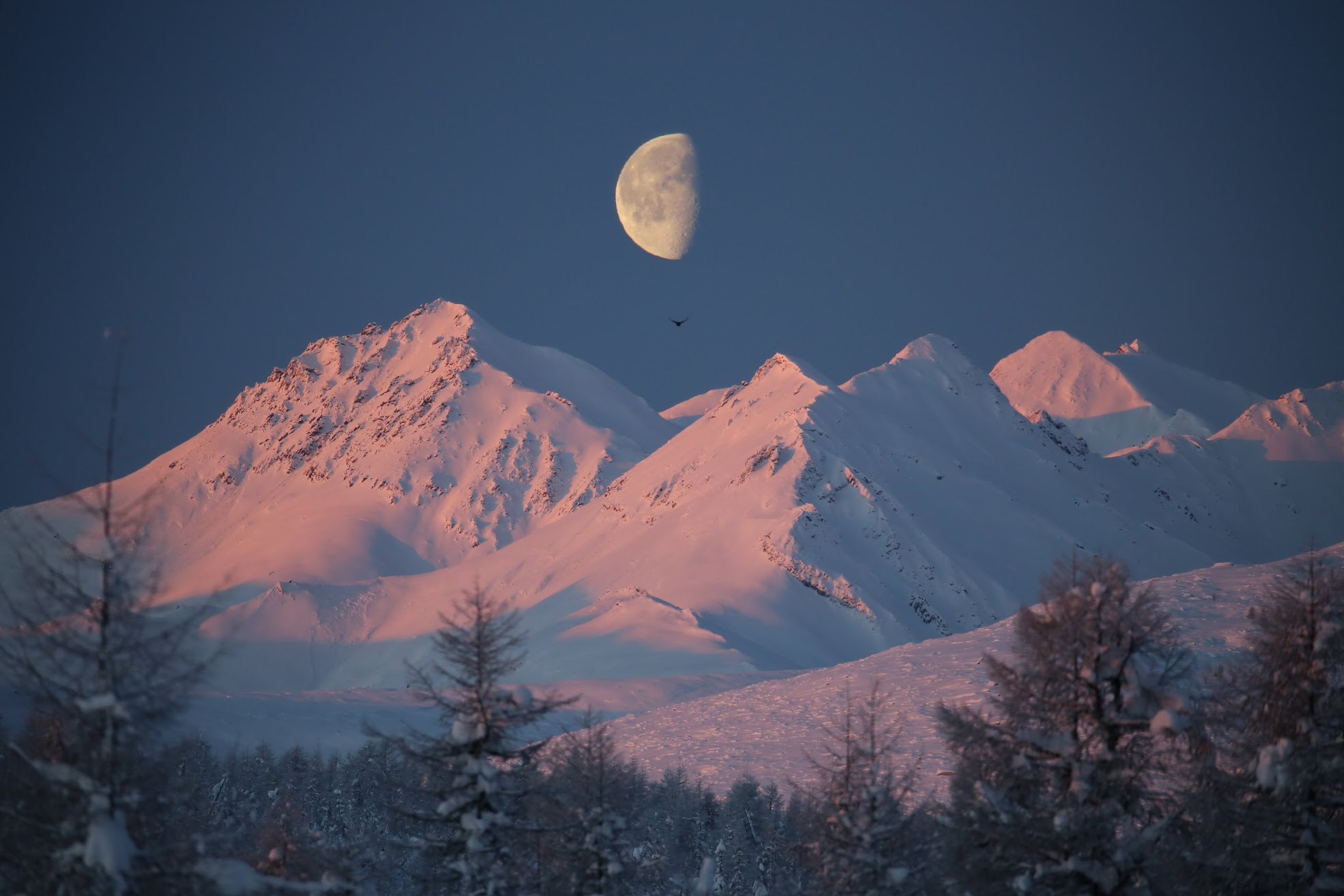}
    \vspace{-3em}
    \caption{JPG quality: 100\%, file size: 823.4 kB}
    \label{fig:jpg-100}
  \end{subfigure}
%
%
  \begin{subfigure}[t]{0.4 \textwidth}
  \captionsetup{labelfont={color=white, scriptsize},font={color=white, scriptsize}}
    \centering
    \includegraphics[width=\textwidth]{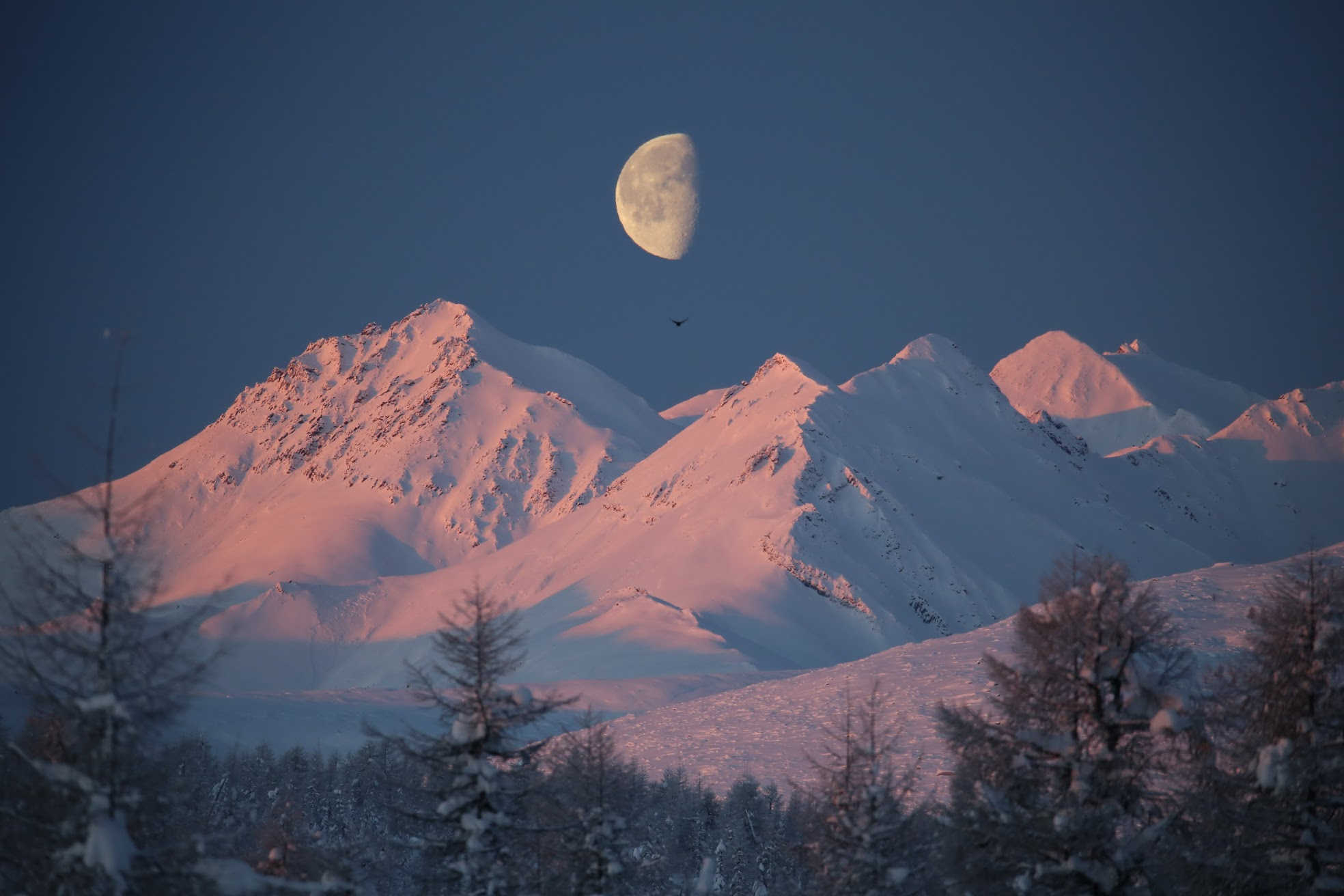}
    \vspace{-3em}
    \caption{JPG quality: 50\%, file size: 130.2 kB}
    \label{fig:jpg-50}
  \end{subfigure}\\
  \vspace{0.2em}
  \begin{subfigure}[t]{0.4 \textwidth}
  \captionsetup{labelfont={color=white, scriptsize},font={color=white, scriptsize}}
    \centering
    \includegraphics[width=\textwidth]{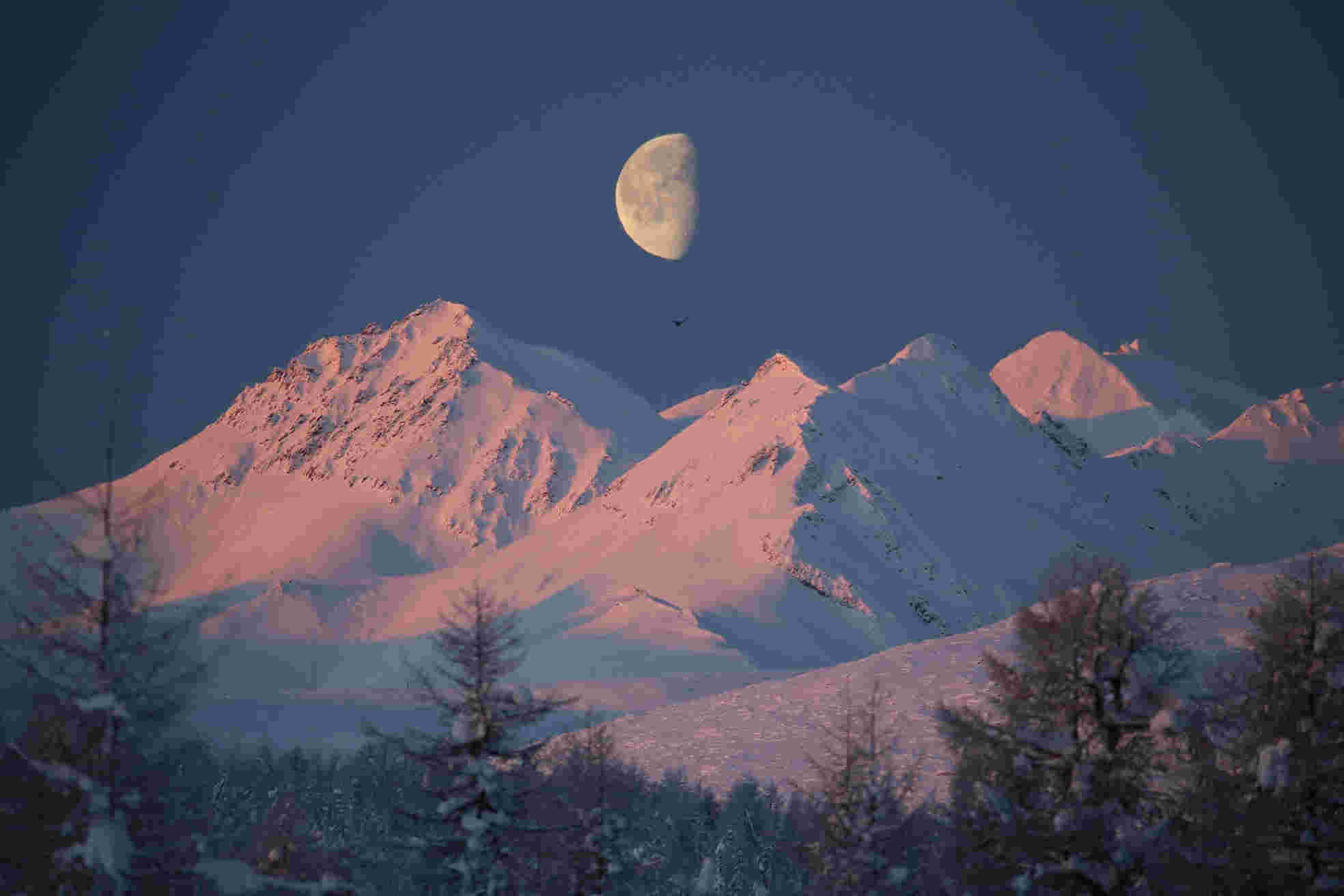}
    \vspace{-3em}
    \caption{JPG quality: 10\%, file size: 50.1 kB}
    \label{fig:jpg-10}
  \end{subfigure}
%
%
  \begin{subfigure}[t]{0.4 \textwidth}
  \captionsetup{labelfont={color=white, scriptsize},font={color=white, scriptsize}}
    \centering
    \includegraphics[width=\textwidth]{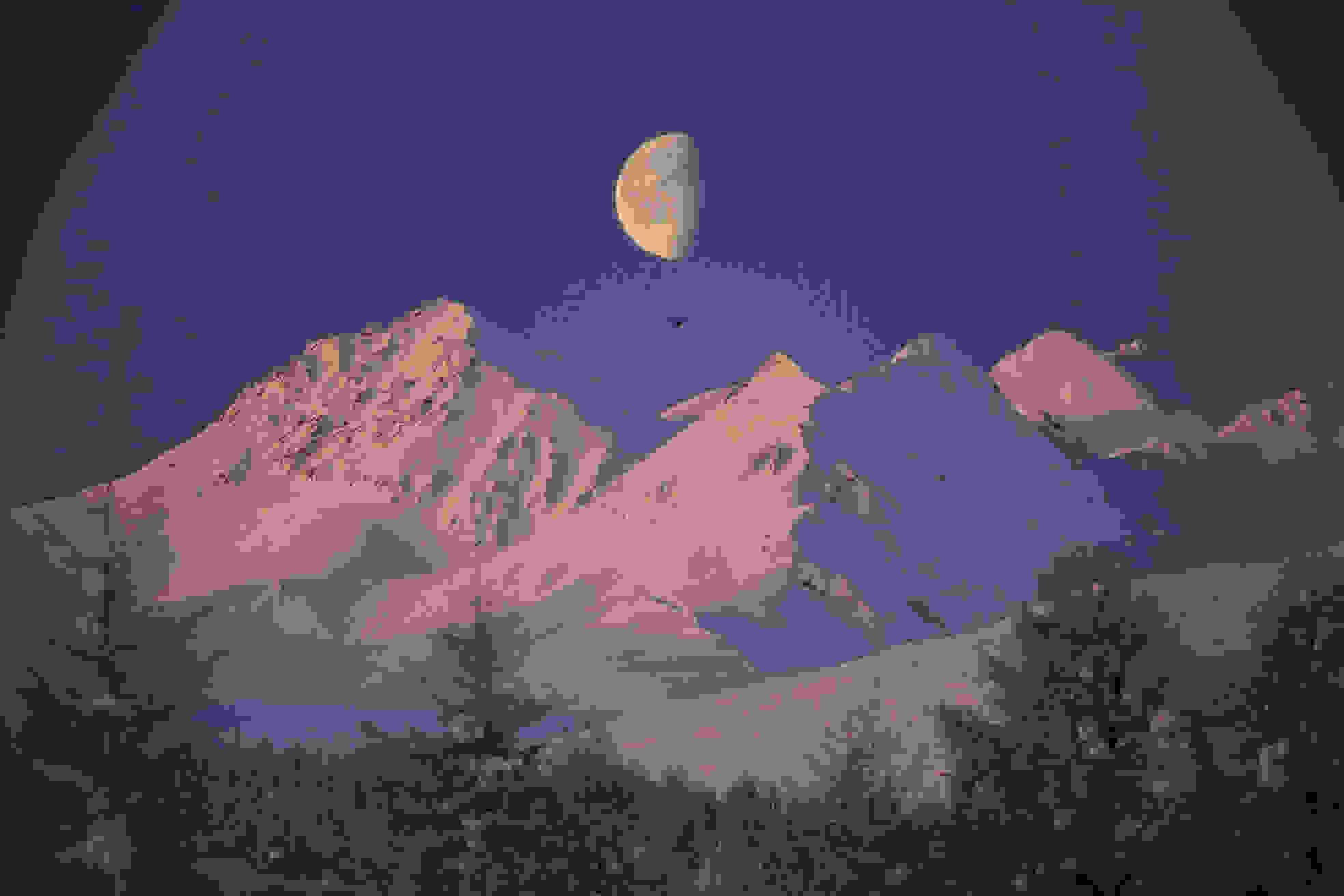}
    \vspace{-3em}
    \caption{JPG quality: 1\%, file size: 33.3 kB}
    \label{fig:jpg-5}
  \end{subfigure}\\
\vspace{2.5em}
\textbf{Lossy graph compression (questions behind Slim Graph):}\\
  \begin{subfigure}[t]{0.8 \textwidth}
    \centering
    \includegraphics[width=\textwidth]{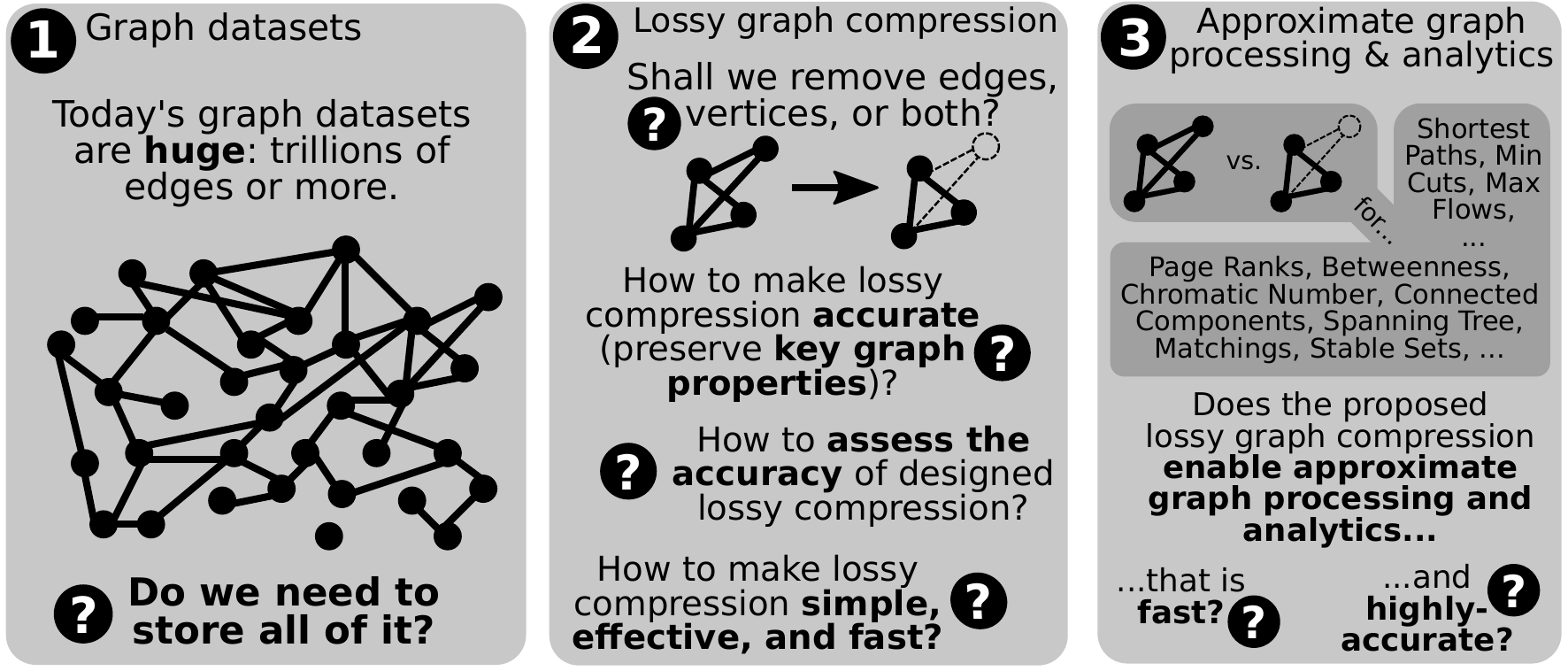}
    \label{fig:lossy-graph-compression-questions}
  \end{subfigure}
%
%
\vspace{-1.5em}
\caption{\textmd{The comparison of different compression levels of the JPG format and
the resulting file sizes 
%
%
%
%
(the photos illustrate Chersky Mountains in Yakutia (North-East Siberia),
in January, with the Moon and an owl caught while flying over taiga forests).
\textbf{Can one apply a similar approach to storing complex graph structures?}}}
%
%
%
%
%
%
\label{fig:jpegs}
\end{figure*}

%
In this work, we argue that the next step towards \emph{significantly} higher
performance and storage reductions in graph analytics can be enabled by
\emph{lossy graph compression} and the resulting \emph{approximate graph
processing}. As the size of graph datasets grows larger,
a question arises: {Does one need to
store and process the exact input graph datasets to ensure precise outcomes of
important graph algorithms?} We show that, as with 
the JPEG compression (see
Figure~\ref{fig:jpegs}),
%
one \emph{may not always} need the full precision 
while processing graphs.



\begin{table*}[t]
\vspace{-1em}
\centering
 \setlength{\tabcolsep}{1.5pt}
\scriptsize
\sf
\begin{tabular}{lllcllll@{}}
\toprule
\multicolumn{3}{l}{\makecell[l]{\textbf{Graph problem + time complexity} \textbf{of the best sequential algorithm}}} & \makecell[c]{\textbf{Output}\\\textbf{type}$^*$} & \makecell[l]{\textbf{Associated graph}\\\textbf{or vertex property}} & \multicolumn{2}{l}{\makecell[l]{\textbf{Associated example parallel algorithm}\\\textbf{(possibly a heuristic) + its time and work}}} & \textbf{Selected application} \\
\midrule
\multirow{5}{*}{\begin{turn}{90}\textbf{Distance}\end{turn}}
 & \makecell[l]{Single-Source Shortest Path (SSSP), unweighted} & $O\left(m+n\right)$~\cite{cormen2009introduction} & V & \makecell[l]{Shortest path length} & BFS~\cite{besta2017push} & $O\left(D d + D \log m\right)$, $O\left(m\right)$  & Bipartite testing \\
 & SSSP, weighted~\cite{cormen2009introduction} & \makecell[l]{$O(m + n \log n)$~\cite{fredman1987fibonacci},\\$O(m)$~\cite{thorup1999undirected}} & V & \makecell[l]{Shortest path length} & $\Delta$--Stepping~\cite{meyer2003delta} & \makecell[l]{$O(d \mathcal{W} \log n + \log^2 n)^{\text{\textdagger}}$,\\$O(n + m + d \mathcal{W} \log n)^{\text{\textdagger}}$}  & Robotics, VLSI design \\
%
%
 & \makecell[l]{Diameter Estimation (DE), Radius Estimation (RE)} & $O(n^2 + nm)$~\cite{shun2013ligra} & S & Diameter, radius & Multiple ($K$) BFS runs~\cite{shun2013ligra} & $O\left(K D d + K D\log m\right)$, $O\left(Km\right)$  & Transportation \\
\midrule
\multirow{4}{*}{\begin{turn}{90}\textbf{Connectiv.}\end{turn}}
 & {[Weakly]} Connected Components~\cite{cormen2009introduction} & $O\left(n+m\right)$~\cite{cormen2009introduction} & S, V & \#Connected components & Shiloach-Vishkin~\cite{shiloach1982logn} & $O\left(\log n\right)$, $O\left(m \log n\right)$ & Verifying connectivity \\
%
%
 & $k$-Core Decomposition (CD)~\cite{khaouid2015k} & $O\left(m\right)$~\cite{batagelj2003m} & S, V & $k$-core-number & Ligra kernel~\cite{shun2013ligra, dhulipala2017julienne} & $O\left(m \log n\right)^{\text{\textdagger}}$, $O\left(m+n\right)^{\text{\textdagger}}$  & Analysis of proteins \\
%
%
 & Triangle Counting (TC)~\cite{shun2015multicore} & $O\left(m d\right)$, $O(m^{3/2})$~\cite{schank2007algorithmic} & S & \#Triangles & GAPBS kernel~\cite{beamer2015gap} & $O(d^2)$, $O\left(m d\right)$  & Cluster analysis \\
%
%
 & \makecell[l]{Low-Diameter Decomposition (LDD)~\cite{miller2015improved}} & $O\left(m\right)$~\cite{miller2015improved} & S, V & --- & Miller et al.~\cite{miller2015improved} & $O\left(\log n \log^* n\right)$, $O\left(m\right)$   & Distance oracles \\
%
\midrule
\multirow{5}{*}{\begin{turn}{90}\textbf{Optimization}\end{turn}}
%
%
& Minimum Spanning Tree (MST)~\cite{cormen2009introduction} & \makecell[l]{$O(m \alpha(m,n))$~\cite{chazelle2000minimum}} & S, V & MST weight & Boruvka~\cite{boruuvka1926jistem} & $O\left(\log n\right)$, $O\left(m \log n\right)$  & Design of networks \\
& Maximum Weighted Matching (MWM)~\cite{papadimitriou1998combinatorial} & $O(m n^2)$ & S, V & MWM weight & Blossom Algorithm~\cite{edmonds1965paths} & --- & Comp. chemistry \\
& Balanced Minimum Cut (MC)~\cite{cormen2009introduction} & NP-Complete & S, V & MC size & METIS kernel~\cite{karypis1995metis} & --- & Network robustness \\
 & Maximum Independent Set (MIS)~\cite{papadimitriou1998combinatorial} & NP-Hard & S, V & MIS size & Luby~\cite{luby1986simple} & $O\left(\log n\right)^{\text{\textdagger}}$, $O\left(m\log n\right)^{\text{\textdagger}}$ & Scheduling \\
%
%
%
%
 & Minimum Graph Coloring (MGC)~\cite{papadimitriou1998combinatorial} & NP-Hard & S, V  & Chromatic number & Jones and Plassmann~\cite{jones1993parallel} & $O\left(\log n / \log \log n\right)^{\text{\textdagger}}$, --- & Scheduling \\
\midrule
\multirow{5}{*}{\begin{turn}{90}\textbf{Central.}\end{turn}}
 & \makecell[l]{Betweenness Centrality (BC)~\cite{brandes2001faster}} & $O\left(n m\right)$, $O(n m +n^2 \log n)^{\text{\textdagger}}$ & V & Betweenness & Brandes~\cite{brandes2001faster} & $O\left(n D d + n D \log m\right)$, $O\left(nm\right)$  & Network analysis \\
%
%
%
%
 & Triangle Counting per Vertex (TCV)~\cite{shun2015multicore} & $O(n d^2), O(m^{3/2})$ & V & \#Triangles & GAPBS kernel~\cite{beamer2015gap} & $O(d^2)$, $O\left(m d\right)$  & Cluster analysis \\
%
%
 & Degree Centrality (DC)~\cite{cormen2009introduction} & $O\left(n+m\right)$ & V, P & Degree & Simple listing~\cite{cormen2009introduction} & $O\left(1\right)$, $O\left(m+n\right)$  & Ranking vertices \\
%
%
 & PageRank (PR)~\cite{page1999pagerank} & $O\left(I m\right)$, $I$ is \#iterations & V, P & Rank & GAPBS kernel~\cite{beamer2015gap} & $O\left(I d\right)$, $O\left(I m\right)$  & Ranking websites \\
\bottomrule
\end{tabular}
%
%
\caption{\textmd{\textbf{Selected considered graph
problems and algorithms.} 
$^*$``V'': a vector of numbers, ``S'': a scalar value, ``P'': a probability distribution.
$^{\text{\textdagger}}$Bounds in expectation or with
high probability.
``---'' means ``unspecified''.
$\mathcal{W}$ is the maximum shortest path weight between any two vertices. 
$K$
is the number of BFS runs in Radii Estimation. 
$I$ is the number of iterations
in PageRank.
$\alpha$ is the inverse Ackermann function.
%
} 
%
%
} 
%
%
\label{tab:problems}
\vspace{-2em}
\end{table*}


Our analogy between compressing graphs and bitmaps brings more 
questions. First, {what is the criterion (or criteria?) of the accuracy
of lossy graph compression?} It is no longer a simple visual similarity as with
bitmaps. Next, {what is the actual method of lossy compression} that
combines large \emph{storage reductions}, high \emph{accuracy}, and \emph{speedups} in graph algorithms
running over compressed datasets?
Finally, {how to easily implement compression schemes?} To answer these
questions, we develop \textbf{\emph{Slim Graph: the first programming model and framework for
lossy graph compression}}. 

The first core idea and element of Slim Graph is a programming model that
enables straightforward development of different compression schemes for
graphs. Here, a developer constructs a simple program called a
\emph{compression kernel}. A compression kernel is similar to a vertex program
in systems such as Pregel~\cite{Malewicz:2010:PSL:1807167.1807184} or
Galois~\cite{nguyen2013lightweight} in that it enables accessing local graph
elements, such as neighbors of a given vertex. However, there are two key
differences. First, the scope of a single kernel is more general than a single
vertex --- it can be an edge, a triangle, or even an arbitrary subgraph.
Second, the goal of a compression kernel is to \emph{remove certain elements of
a graph}. The exact elements to be removed are determined by the body of a
kernel. In this work, we introduce kernels that preserve graph properties as
different as Shortest Paths or Coloring Number while removing significant
fractions of edges; these kernels constitute \emph{novel graph compression
schemes}.  We also illustrate kernels that implement
{spanners}~\cite{peleg1989graph} and {spectral
sparsifiers}~\cite{spielman2011spectral}, established structures in graph
theory. These are graphs with edges removed in such a way that, respectively,
the distances between vertices and the graph spectra are preserved up to
certain bounds.  Finally, for completeness, we also express and implement a
recent variant of {lossy graph summarization}~\cite{shin2019sweg}.
\emph{Based on the analysis of more than 500 papers on graph compression, we
conclude that \textbf{Slim Graph enables expressing and implementing all major
classes of lossy graph compression}}, including sampling, spectral sparsifiers,
spanners, graph summarization, and others.

\iftr
We 
illustrate
that they can be used to preserve practically
important graph properties. 
\fi

\ifsc
\fi

Next, Slim Graph contributes \emph{metrics} for
assessing the accuracy of lossy graph compression. 
\iftr
For algorithms that output
scalar numbers (e.g., Boruvka Algorithm for computing a Minimum Spanning
Tree~\cite{boruuvka1926jistem}), we use a simple relative error (e.g., the
change in the tree weight). 
\fi
For algorithms that assign certain values to each
vertex or edge that impose some vertex or edge ordering (e.g., Brandes
Algorithm for Betweenness Centrality~\cite{brandes2001faster}), we analyze the
numbers of vertex or edge pairs that switched their location in the order after
applying compression. Moreover, for graph algorithms with output that can be
interpreted as a probability distribution (e.g.,
PageRank~\cite{page1999pagerank}), we propose to use \emph{statistical
divergences}, a powerful tool used in statistics to assess the similarity and
difference of two probability distributions. We analyze a large number of
difference divergence measures and we select the \emph{Kullback-Leibler
divergence}~\cite{kldivergence} as the most suitable tool in the context of 
comparing graph structure.

\iftr
We also analyze and compare Slim Graph to the few existing graph lossy
compression schemes, such as
%
%
cut sparsifiers~\cite{benczur1996approximating}, low-rank
approximation~\cite{savas2011clustered}, lossy graph
summarization~\cite{navlakha2008graph}, and
others~\cite{tsourakakis2011triangle, liu2012compressing}. We show that these
methods are fundamentally not scalable.
\fi

We conduct a \emph{theoretical analysis}, presenting or deriving \emph{more than
50 bounds} that illustrate how graph properties change under different
compression methods.
We also {evaluate} Slim Graph for different algorithms, on both shared-memory
high-end servers and distributed supercomputers.
Among others, we were able to use Slim Graph to compress Web Data Commons 2012,
the largest publicly available graph that we were able to find (with
$\approx$3.5 billion vertices and $\approx$128 billion edges), reducing its
size by 30-70\% using distributed compression.
%
%
{\emph{Slim Graph
may become a common ground for developing, executing,
and analyzing emerging lossy graph compression schemes
on shared- and distributed-memory systems}}.

%

\begin{figure*}[t]
\centering
\includegraphics[width=0.7\textwidth]{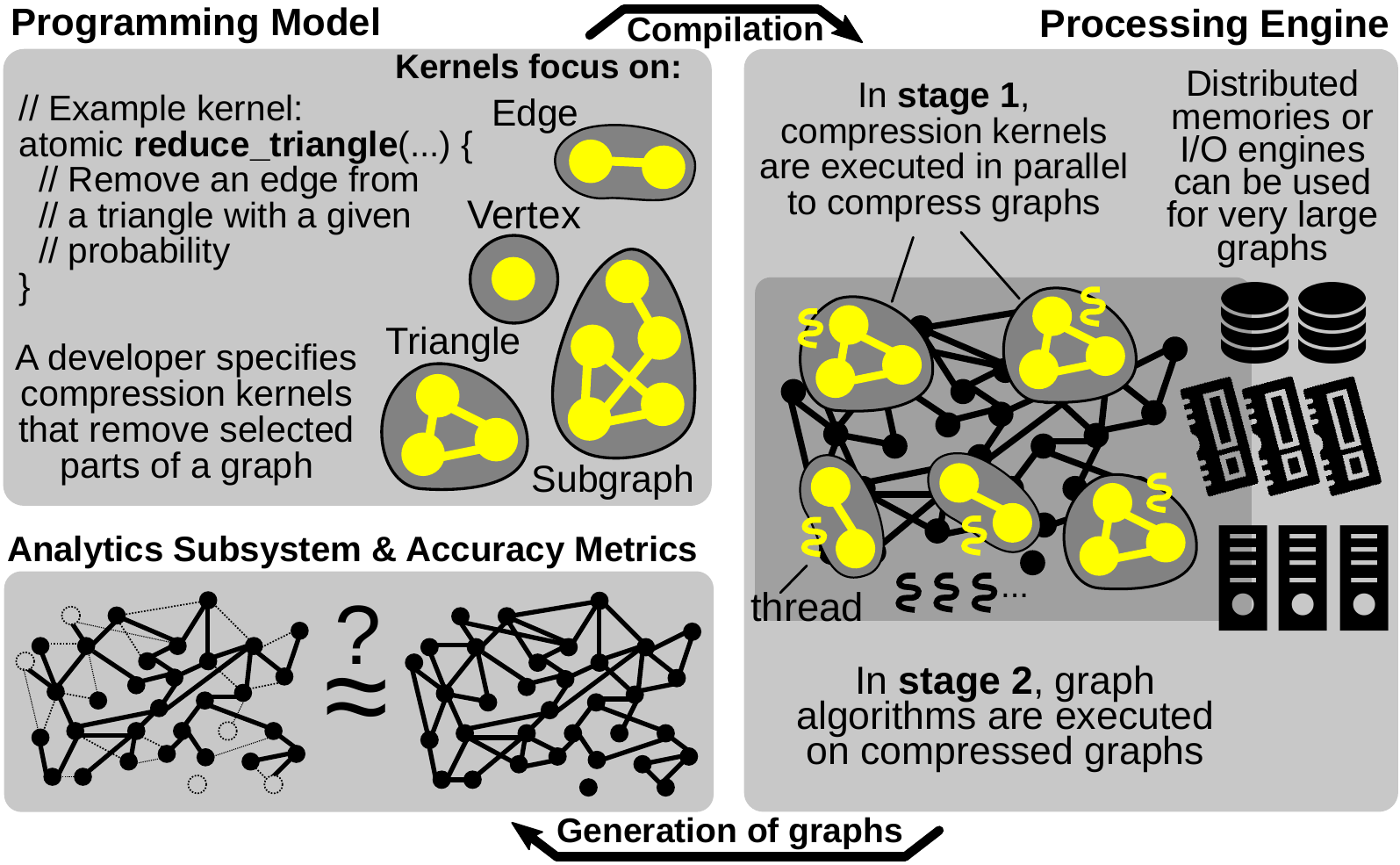}
%
\caption{\textmd{The overview of the architecture of Slim Graph.}}
\label{fig:engine-overview}
%
\end{figure*}

\vspace{0.25em}
\noindent
To summarize, we offer the following contributions:
\vspace{-0.25em}

\begin{itemize}[noitemsep, leftmargin=1em]
\item We introduce Slim Graph: \textbf{the first programming model and framework} 
for \textbf{practical and efficient lossy graph compression}. The Slim Graph programming
model is based on a \textbf{novel abstraction of compression kernels} that is versatile:
it enables expressing different graph compression and sparsification methods such as
simple random uniform sampling, spanners, or spectral sparsifiers.
%
%
\item We propose \textbf{Triangle Reduction}: a scalable \textbf{method for lossy graph
compression} that preserves well various graph properties such as connectivity or distances.
\item We establish a \textbf{set of metrics} for comparing the accuracy of
lossy graph compression schemes for many graph problems and algorithms. Our
metrics range from simple relative changes in scalar numbers to {statistical
tools such as the Kullback-Leibler divergence}.
\item We conduct a \textbf{theoretical analysis} of Slim Graph and illustrate
how its different routines impact 12 graph properties.
\item We offer a \textbf{supporting taxonomy} of the most important classes of
graph sparsification and lossy compression approaches, and a set of \textbf{guidelines}
on when to use which of the considered compression algorithms and routines.
\item We conduct an extensive \textbf{large-scale evaluation}, demonstrating the advantages of Slim Graph over
past lossy compression schemes. We offer the first results of
distributed graph compression, and we show that
\textbf{Slim Graph enables highly-accurate and high-performance approximate
graph processing, storage, and analytics}. 
\end{itemize}



\section{Notation And Background}
\label{sec:back}

We first summarize basic concepts. 
\ifscr
Tables~\ref{tab:abbr} and~\ref{tab:symbols} present the used abbreviations and symbols, respectively.
\fi

\begin{table}[h]
\centering
 \setlength{\tabcolsep}{1pt}
\footnotesize
\small
\sf
\begin{tabular}{@{}ll@{}}
\toprule
BFS, SSSP & Breadth-First Search, Single Source Shortest Path~\cite{cormen2009introduction}\\
MST, PR, CC & Min.~Spanning Tree, PageRank~\cite{page1999pagerank}, Conn.~Components\\ 
BC, TC & Betweenness Centrality~\cite{brandes2001faster, solomonik2017scaling}, Triangle Counting~\cite{shun2015multicore}\\
\midrule
TR, EO, CT & Triangle Reduction, Edge Once, Count Triangles\\
KL, SVD & Kullback-Leibler, Singular Value Decomposition\\
SG & Slim Graph\\
%
\bottomrule
\end{tabular}
\caption{The most important \textbf{abbreviations used in the paper}.}
\label{tab:abbr}
\vspace{-1em}
\end{table}

\begin{table}[h]
\centering
 \setlength{\tabcolsep}{2pt}
\small
\sf
\begin{tabular}{@{}ll@{}}
\toprule
%
%
       $G$&A graph $G=(V,E)$; $V$ and $E$ are sets of vertices and edges.\\
       $n,m$&Numbers of vertices and edges in $G$; $|V| = n, |E| = m$.\\
       $d_v, N_v$ & The degree and neighbors of a vertex $v$. \\
       $d$ & The maximum vertex degree.\\
       $D$ & The diameter of a given graph.\\
       $T$ & The total number of triangles in a given graph.\\
%
%
%
       $dist_G(v,u)$ & The shortest path length between vertices $v,u$ in graph~$G$.\\
       $I$ & The number of iterations, details depend on the context.\\
       $k$ & An input parameter, details depend on the context. \\
       $\epsilon$ & An accuracy parameter, details depend on the context. \\
%
%
%
%
\midrule
%
%
       $T,P$ & The number of threads/processes.\\
       $W$ & The memory word size [bits].\\
\bottomrule
\end{tabular}
\caption{The most important \textbf{symbols used in the paper}.}
\label{tab:symbols}
\end{table}


%
We \macb{model} an undirected graph $G$ as a tuple $(V,E)$; $V$ is a set of
vertices and $E \subseteq V \times V$ is a set of edges; $|V|=n$, $|E|=m$.
$N_v$ and $d_v$ denote the neighbors and the degree of a vertex $v$, respectively.
We also consider weighted and directed graphs and mention this appropriately.
\iftr
For weighted graphs, $W: E \to \mathbb{R}$
assigns weights to edges; $W(v,w) \equiv W(e)$ is the weight of an edge~$e =
(v,w)$.
For directed graphs, $E$ becomes a set of \emph{arcs}.
\fi
\iftr
For directed graphs, we also use the notions of $v$'s in-degree~$d_{i,v}$,
out-degree~$d_{o,v}$, in-neighbors~$N_{i,v}$, and out-neighbors~$N_{o,v}$.
\fi
The shortest path length between vertices $u$ and $v$ in a graph $G$ is
$dist_G(u,v)$. $G$'s maximal degree and diameter are ${d}$
and $D$. $T$ is the total number of triangles in a
graph. 
\iftr
$T_V \equiv T / n$ and $T_E \equiv T / m$ are the average number of triangles
per vertex and per edge.
\fi
%

%
\ifsc
We list considered \macb{graph problems}
in Table~\ref{tab:problems}.
They can be loosely grouped into
problems related to vertex distances, connectivity, optimization, and 
centrality measures.
\fi
\iftr
We now list considered well-known \macb{graph problems}.
These are BFS~\cite{cormen2009introduction}, Single-Source Shortest Path~\cite{cormen2009introduction},
Connected Components (CC)~\cite{cormen2009introduction},
Triangle Counting (TC)~\cite{shun2015multicore},
Minimum Spanning Tree (MST)~\cite{cormen2009introduction},
Maximum Matching~\cite{papadimitriou1998combinatorial},
Maximum Independent Set~\cite{papadimitriou1998combinatorial},
Minimum Graph Coloring~\cite{papadimitriou1998combinatorial},
Betweenness Centrality (BC)~\cite{brandes2001faster},
and PageRank (PR)~\cite{page1999pagerank}.
\fi
\iftr
and PageRank (PR)~\cite{page1999pagerank};
selected details are
presented in Table~\ref{tab:problems}. 
\fi
Importantly, the complexity of algorithms 
solving these problems is proportional to $m$.
\iftr
($m$ and $d$).
\fi
\emph{Thus, removing graph edges would directly accelerate
the considered graph algorithms.}

\iftr
\subsection{Details of Considered Graph Problems}
\noindent \ding{182}
\macb{Distance Problems} 
First, we consider \emph{pairwise distances} between
vertices. 
%
%
In \textbf{Single Source Shortest Paths (SSSP)}~\cite{cormen2009introduction},
separately for weighted and unweighted graphs, we consider the distance from
each vertex to a selected root vertex, both for unweighted and weighted edges.
In \textbf{Radii Estimation (RE)}~\cite{shun2013ligra}, we approximate the
radius of each vertex~$v$: the shortest distance from $v$ to the furthest
reachable vertex.

\noindent \ding{183}
\macb{Connectivity Problems}
Second, we inspect graph \emph{connectivity} properties, most importantly the
number of graph \textbf{Connected Components
(CC)}~\cite{cormen2009introduction}, also referred to as \textbf{Weakly
Connected Components (WCC)}, (subgraphs in which any two vertices are connected
with an undirected path).
%
%
Moreover, we consider the \textbf{$k$-Core Decomposition
(CD)}~\cite{khaouid2015k}: a well-established metric that, intuitively,
partitions a graph into layers from ``central'' to ``peripheral'' vertices.
Another measure is \textbf{Triangle Counting (TC)}~\cite{shun2015multicore} in
which one counts the number of triangles in a graph. Finally, we consider the
\textbf{Low-Diameter Decomposition (LDD)}~\cite{miller2015improved} in which a
graph is decomposed into subgraphs with low diameters and few edges connecting
different subgraphs.

\noindent \ding{184}
\macb{Optimization Problems}
Another class of problems is related to minimizing or maximizing a certain
graph property. Here, we analyze the derivation of the \textbf{Minimum Spanning
Tree (MST)}~\cite{cormen2009introduction} (a spanning tree of an undirected
graph with the smallest possible sum of weights of the tree edges), the
\textbf{Maximum Flow (MF)}~\cite{cormen2009introduction} (the maximum flow in a
directed graph), the \textbf{Minimum Cut (MC)}~\cite{gianinazzi2018communication, geissmann2018parallel},
the \textbf{Maximum Independent Set (MIS)}~\cite{papadimitriou1998combinatorial}
(the largest set of vertices such that no two vertices are adjacent),
the \textbf{Maximum Weighted Matching (MWM)}~\cite{papadimitriou1998combinatorial}
in general graphs (the largest set of edges such that no two edges are adjacent
to the same vertex), and the \textbf{Maximum Weighted Bipartite Matching
(MWBM)}~\cite{papadimitriou1998combinatorial} in bipartite graphs. We also
consider the \textbf{Minimum Graph Coloring
(MGC)}~\cite{papadimitriou1998combinatorial}, a problem that aims at minimizing
the number of labels assigned to vertices such that no two adjacent vertices
have the same label.

\noindent \ding{185}
\macb{Centrality Problems}
We consider several measures of vertex centrality. In
\textbf{Betweenness Centrality (BC)}~\cite{brandes2001faster}, one derives the
centrality of each vertex~$v$ based on the number of shortest paths 
between any other pairs of vertices, going via $v$. In \textbf{Triangle Counting per
Vertex (TCV)}~\cite{shun2015multicore}, the importance of a vertex is based on
the number of triangles that a vertex belongs to.  \textbf{Degree Centrality
(DC)} is a simple count of neighbors of each vertex.  Finally, we consider
\textbf{PageRank (PR)}~\cite{page1999pagerank}, the well-known centrality
measure where the importance of a vertex~$v$ is based on the number of other
important vertices being adjacent to $v$.
\fi

\iftr
\subsection{Naming of Graph Compression}
\fi
We also clarify \macb{naming}: we use the term ``lossy graph compression'' to
refer to any scheme that removes some parts of graphs: \emph{sparsification}
and \emph{sparsifiers}~\cite{benczur1996approximating, spielman2011spectral},
\emph{sketches}~\cite{ahn2012analyzing}, \emph{synopses}~\cite{guha2012graph},
\emph{sampling}~\cite{hu2013survey, wang2011understanding,
leskovec2006sampling}, \emph{spanners}~\cite{peleg1989graph}, \emph{low-rank
approximation}~\cite{savas2011clustered, sui2012parallel}, \emph{bounded-error
summarization}~\cite{navlakha2008graph}, \emph{lossy
compression}~\cite{henecka2015lossy}, and \emph{reduction}~[This work].

\section{Slim Graph Architecture}
\label{sec:schemes}

We now describe the architecture of Slim Graph. An overview is presented
in Figure~\ref{fig:engine-overview}. Slim Graph consists of three key
elements: (1) a programming model, (2) an execution engine, and (3)
an analytics subsystem with accuracy metrics.

\subsection{Part One: Programming Model}
 
The first core part of Slim Graph is a \emph{programming model for graph
compression}. The model provides a developer with a set of {programmable
compression kernels} that  can be used to \emph{express and implement graph compression schemes}.
%
%
Intuitively, the developer can program the kernel by providing a small code
snippet that uses the information on graph structure (provided by the kernel
arguments) to remove certain parts of the graph.  These kernels are then
executed by an underlying engine, where multiple \emph{instances of kernels}
run in parallel.

Thus, a developer has a ``local'' view of the input graph~\cite{tate2014programming}, similar to that of
vertex-centric processing frameworks such as
Pregel~\cite{Malewicz:2010:PSL:1807167.1807184} or
Galois~\cite{nguyen2013lightweight}. Still, Slim Graph
enables \emph{several} types of kernels where the ``local view of the graph'' is 
(1) a vertex and its neighbors, but it can also be (2) an edge with adjacent vertices,
(3) a triangle with neighboring vertices, and (4) a subgraph with a list of
pointers to vertices within the subgraph and pointers to neighboring
vertices. As we show in detail in~\cref{sec:schemes}, \emph{each type of
kernel is associated with a certain class of graph compression algorithms}.
For example, a subgraph is used to implement spanners while a triangle is
associated with Triangle Reduction, a class proposed in this work. Each of
these classes can be used to reduce the graph size while preserving
\emph{different} properties; we provide more details in~\cref{sec:theory}
and~\cref{sec:evaluation}.
\emph{Slim Graph offers multiple compression schemes because no single
compression method can be used to preserve many graph properties deemed
important in today's graph computations.}

The developer can indicate whether
different parts of a compression kernel will execute \emph{atomically}~\cite{schweizer2015evaluating}. The
developer can also specify if a given element should be considered for removal
\emph{only once} or \emph{more than once} (i.e., by more than one kernel instance).
This enables various tradeoffs between
performance, scope (i.e., number of removed graph elements), and accuracy of compression.
%

For example, while iterating over triangles and removing one triangle edge with
a certain probability~$p$, one can ensure that this edge -- assuming it is not
removed in one kernel instance -- will not be considered for removal again in
other instances.
Thus, edges that belong to multiple triangles (and are potentially more important by,
e.g., being a part of several shortest paths) are not removed 
more often than edges that are a part of one triangle. We discuss this and
other such schemes in more detail in~\cref{sec:schemes}.


\begin{table*}[t]
\centering
 \setlength{\tabcolsep}{2pt}
  \renewcommand{\arraystretch}{1.2}
\footnotesize
\sf
\begin{tabular}{llllll@{}}
\toprule
\makecell[c]{\textbf{Compression scheme}} & \makecell[c]{\textbf{\#remaining edges}} & \makecell[c]{\textbf{Work}} & \makecell[c]{\textbf{W, D}$^\text{\textdagger}$} & \makecell[c]{\textbf{Storage}}$^\text{\textdollar}$ & \textbf{Preserves best...} \\
\midrule
\multicolumn{6}{c}{\makecell[c]{Lossy compression schemes that are a \textbf{part of Slim Graph}.}} \\
\midrule
%
%
\makecell[l]{\textbf{(\cref{sec:kernel-spectral}) Spectral sparsification} (``High-conductance'' sampling~\cite{spielman2011spectral})} & $\propto\max(\log \frac{3}{p}, \log n) n$ & $m + O(1)$ & \faThumbsDown, \faThumbsOUp & $O(m+n)$ & Graph spectra \\
\textbf{(\cref{sec:sg-ssp}) Edge sampling} (simple random-uniform sampling) & $(1-p)m$ & $m + O(1)$ & \faThumbsDown, \faThumbsOUp & $O(m+n)$ & Triangle count \\
\textbf{(\cref{sec:sg-tr}) Triangle reduction} (several variants are described in~\cref{sec:sg-tr}) & $m - p T$ (more in~\cref{sec:theory}) & $O(n d^2)$ or $O(m^{3/2})$ & \faThumbsOUp, \faThumbsOUp & $O(m+n)$ & Several (\cref{sec:theory}) \\
\textbf{(\cref{sec:sg-spanners}) Spanners} ($O(k)$--spanner~\cite{miller2015improved}) & \makecell[l]{$O(n^{1+{1}/{k}} \log k )$} & $O(m)$ & \faThumbsOUp, \faThumbsDown & $O(m+n)$ & Distances \\
\textbf{(\cref{sec:sg-summaries}) Lossy summarization with Jaccard similarity} (SWeG~\cite{shin2019sweg}) & $m \pm 2\epsilon m ^{\text{\textdaggerdbl}}$ & {$O(m I )^{\text{\textdaggerdbl}}$} & \faThumbsDown, \faThumbsDown$^*$ & $O(m+n)$ & \makecell[l]{Count of common neighbors} \\
%
%
\midrule
\multicolumn{6}{c}{\textbf{Past schemes} for lossy graph compression (some might be integrated with Slim Graph in future versions):} \\
\midrule
\textbf{(\cref{sec:sg-ot}) Lossy summarization with the MDL principle} (ApxMdl~\cite{navlakha2008graph}) & $\epsilon m ^{\text{\textdaggerdbl}}$ & {$O(C^2 \log n + n m_S)^{\text{\textdaggerdbl}}$} & \faThumbsDown, \faThumbsOUp & $O(m+n)$ & Unknown \\
\textbf{(\cref{sec:sg-ot}) Lossy linearization}~\cite{maserrat2012community} & $2 k n ^*$ & $O(m d I T)^*$ & \faThumbsDown, \faThumbsDown & $O(m+n)$ & Unknown \\
\textbf{(\cref{sec:sg-ot}) Low-rank approximation} (clustered SVD~\cite{sui2012parallel, savas2011clustered}) & --- & {$O(n_c^3)^{\text{\textdaggerdbl}}$} & \faThumbsOUp, \faThumbsOUp & $O(n_{c}^2)^{\textdaggerdbl}$ & [High error rates] \\
\textbf{(\cref{sec:sg-ot}) Cut sparsification} (Benczúr--Karger~\cite{benczur1996approximating}) & $O(n \log n \ \epsilon^2)$ & {$O(m \log^3 n +m \log n / \epsilon^2)^{\text{\textdaggerdbl}}$} & \faThumbsOUp, \faThumbsDown & $O(n+m)$ & Cut sizes \\
\bottomrule
\end{tabular}
%
\caption{\textmd{(\cref{sec:schemes}) \textbf{Considered lossy compression schemes.}
\iftr
A detailed list is in a full 
survey on lossy graph compression that will be released upon the publication of Slim Graph~\cite{koutis2016simple, elkin2018efficient, chu2018graph,
lenzen2018centralized, censor2018sparsest, censor2018distributed,
dory2018distributed, lenzen2018centralized, parter2018local, kyng2017framework,
elkin2016distributed, alstrup2017constructing, parchas2018uncertain,
durfee2017sampling, kyng2018matrix, calandriello2018improved,
jambulapati2018efficient, alev2017graph, goranci2017power, liu2019short,
li2018spectral, soma2019spectral, carlson2019optimal, saranurak2019expander,
wang2017towards, anderson2014efficient, feng2016spectral, wang2017towards,
sui2012parallel, savas2016clustered}
\cite{riondato2017graph, liu2018graph, shin2019sweg,
campinas2013efficiency, toivonen2011compression, chen2009mining, liu2014distributed, fan2012query, toivonen2012network, sadri2017shrink, zhou2017summarisation,
tsalouchidou2018scalable, beg2018scalable, kumar2018utility, dunne2013motif, maserrat2012community, kalavri2016shortest}.
\fi
$^{\text{\textdagger}}$\textbf{W,D} indicate support for weighted or
directed graphs, respectively.
Symbols used in Slim Graph schemes ($p,k$) are explained in corresponding sections.
$^{\text{\textdollar}}$Storage needed to conduct compression.
%
%
%
In the \textbf{SWeG lossy summarization}~\cite{shin2019sweg}, $\epsilon$ controls the
approximation ratio
while $I$ is the number of iterations (originally set to 80~\cite{shin2019sweg}).
  $^*$SWeG covers undirected graphs but uses a compression metric for directed graphs.
In \textbf{ApxMdl}~\cite{navlakha2008graph}, $\epsilon$ controls the
approximation ratio, $C \in O(m)$ is the number of ``corrections'', $m_S
\in O(m)$ is the number of ``corrected'' edges.
In \textbf{lossy linearization}~\cite{maserrat2012community},
$k \in O(n)$ is a user parameter, $I$ is the number of iterations of a ``re-allocation process'' (details in Section~V.C.3 in the original work~\cite{maserrat2012community}),
while $T$ is a number of iterations for the overall algorithm convergence.
In \textbf{clustered SVD approximation}~\cite{sui2012parallel, savas2011clustered}, $n_c
\le n$ is the number of vertices in the largest cluster in low-rank
approximation.
%
%
In \textbf{cut sparsifiers}~\cite{benczur1996approximating}, $\epsilon$
controls the approximation ratio of the cuts.}
}
%
\label{tab:theory-schemes}
\end{table*}

\subsection{Part Two: Execution Engine}


Second, Slim Graph's processing engine executes compression
kernels over input graphs, performing the actual compression.  The engine
consists of a two-stage pipeline. In stage~1, a graph is compressed with 
a selected method. In stage~2, a selected graph algorithm is executed on the
compressed graph to verify how compression impacts the graph structure.
Many considered real-world graphs fit in a memory of a single ``fat'' server
and we use established in-memory techniques and integrate Slim Graph with
high-performance shared-memory processing infrastructure, namely GAP Benchmark
Suite~\cite{beamer2015gap}, to deliver fast graph compression routines
(we extend GAPBS with new graph algorithms whenever necessary, e.g., to compute matchings, spanning
trees, and others).
However, if graphs do not fit into the memory of one server, we use a separate
pipeline with I/O and distributed-memory tools. 
Currently, we use
a distributed-memory implementation of edge compression kernels, based on 
MPI Remote Memory Access~\cite{gerstenberger2014enabling, di2019network,
besta2015active, schmid2016high, besta2014fault}.
\iftr
\maciej{fix}
Currently, we use
GraphChi~\cite{kyrola2012graphchi} (an established I/O graph
framework)
and Distributed Galois~\cite{nguyen2013lightweight, dathathri2018gluon}
(an established distributed graph framework).
\fi

Challenges behind designing fast graph processing engines were studied
thoroughly in the last decade and summarized in numerous
works~\cite{besta2017push, lu2014large, doekemeijer2014survey,
heidari2018scalable, besta2019demystifying, besta2017slimsell, besta2019graph, zhang2016parallel, zhang2015memory, shi2018graph,
lee2012parallel, batarfi2015large, mccune2015thinking, besta2015accelerating}.  Thus, in the
following, we focus on the novel contributions, which are (1) kernel abstractions
for graph compression, (2) novel graph compression methods, (3)
novel accuracy metrics, and (4) theoretical and empirical evaluation.




\subsection{Part Three: Analytics Subsystem}

Slim Graph also provides methods and tools for analyzing the accuracy of graph
compression schemes. The proposed metrics can be used to compare the outcomes
of graph algorithms that generate a scalar output (e.g., a number of Connected
Components), a vector (e.g., Betweenness Centrality), or a probability
distribution (e.g., PageRank). The results of this analytics can be used by
the Slim Graph user to provide feedback while implementing graph compression
routines. We discuss these metrics in detail in~\cref{sec:metrics}.

%
%

\ifscr
\begin{figure*}[hbtp]
\centering
\vspace{-2em}
\centering
\includegraphics[width=0.99\textwidth]{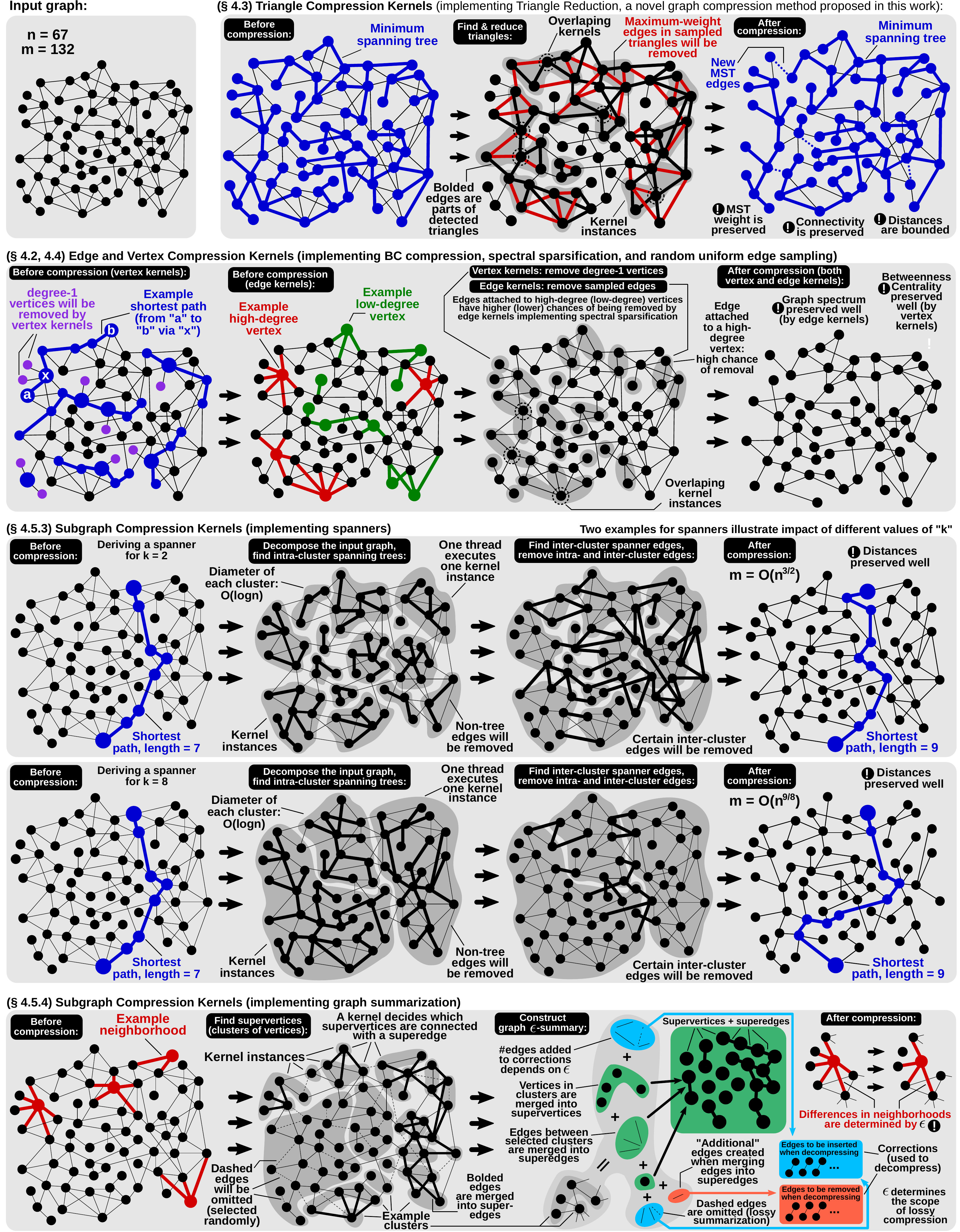}
\centering
\vspace{-1em}
\caption{The overview of Slim Graph compression kernels and their impact on various graph properties. \textmd{Vertex/edge kernels are shown together due to space constraints.}}
\label{fig:slim-graph}
%
\end{figure*}
\fi


\section{Slim Graph: Compressing Graphs}
\label{sec:schemes}


We now show how to \textbf{develop lossy graph compression
schemes using Slim Graph abstraction of compression kernels}.  
Table~\ref{tab:theory-schemes} summarizes schemes
considered in this work.
We (1)
describe each scheme and (2) provide the
pseudocode of the corresponding kernels and any other required structures.
Throughout this section, we use Figure~\ref{fig:slim-graph}
(overview of kernels) and Listing~\ref{lst:kernels}.
In Listing~\ref{lst:kernels}, we show the ``C++ style'' pseudocode, focusing on
kernels as the most complex and compression-related part of Slim Graph. 
We present the code of seven key kernels; more examples (a
total of 16) can be found in the extended technical report.
Finally, we propose \textbf{Triangle Reduction, a tunable class of
graph compression schemes}, together with corresponding kernels.




\subsection{Compression Kernels: Syntax + Semantics}
\label{sec:syntax-semantics}

We summarize selected parts of Slim Graph syntax and semantics.
%
  %
To implement a kernel, one first specifies a kernel name and a single kernel argument
\texttt{x}; \texttt{x} can be a vertex, an edge, a triangle, or a subgraph.
Within kernel's body, \texttt{x} offers properties and methods that enable
accessing and modifying local graph structure, e.g., edges adjacent to \texttt{x}.
\iftr
\maciej{fix}
When using subgraphs, one also determines a mapping \texttt{m} between vertices
and subgraphs.
\fi
Slim Graph also provides a global container object \texttt{SG}. \texttt{SG}
offers various functions and parameters for accessing or modifying
\emph{global} graph structure, for example \texttt{del(a)} (delete a graph
element \texttt{a})
%
%
or \texttt{out\_edges(X)} (return all edges with a source vertex in a subgraph
induced by elements \texttt{X}). \texttt{SG} also contains properties of the
used compression scheme, for example values of sampling parameters.  Finally,
Slim Graph syntax includes a keyword \texttt{atomic} (it indicates atomic
execution) and opaque reference types for vertices and edges (\texttt{V} and
\texttt{E}, respectively).  Example \texttt{V} fields are \texttt{deg} (degree)
and \texttt{parent\_ID} (ID of the containing graph element, e.g., a subgraph).
Example \texttt{E} fields are \texttt{u} (source vertex), \texttt{v}
(destination vertex), and \texttt{weight}.

\iftr
Slim Graph utilizes system annotations to mark atomic regions. 
Moreover, we provide a global singleton object \texttt{SG} with various
properties such as $n$, $m$.
\fi

\begin{lstlisting}[aboveskip=1em,abovecaptionskip=-0.1em,float=t,label=lst:kernels,caption=\textmd{Lossy graph compression schemes with
Slim Graph.}]
|\vspace{0.25em}|/********** |\textbf{\ul{Single-edge compression kernels}}| (|\cref{sec:sg-ss}|) ****************/
|\label{ln:spectral-1}|spectral_sparsify(E e) { //More details in |\cref{sec:kernel-spectral}|
  double $\Upsilon$ = SG.connectivity_spectral_parameter();
  double edge_stays = min(1.0, $\Upsilon$ / min(e.u.deg, e.v.deg));
  if(edge_stays < SG.rand(0,1)) 
    atomic SG.del(e);
|\label{ln:spectral-2}|  else e.weight = 1/edge_stays; 
|\vspace{0.5em}|}

|\label{ln:uniform-1}|random_uniform(E e) { //More details in |\cref{sec:sg-ssp}|
  double edge_stays = SG.$p$;
  if(edge_stays < SG.rand(0,1)) 
|\label{ln:uniform-2}|    atomic SG.del(e); 
|\vspace{0.5em}|}

|\vspace{0.25em}|/************* |\textbf{\ul{Triangle compression kernels}}| (|\cref{sec:sg-tr}|) ****************/
|\label{ln:triangle-1}|p-1-reduction(vector<E> triangle) { 
  double tr_stays = SG.$p$; 
  if(tr_stays < SG.rand(0,1))  
|\label{ln:triangle-2}|    atomic SG.del(rand(triangle)); 
|\vspace{0.5em}|}

|\label{ln:triangle-once-1}|p-1-reduction-EO(vector<E> triangle) { 
  double tr_stays = SG.$p$; 
  if(tr_stays < SG.rand(0,1)) {
    E e = rand(triangle); 
    atomic {
      if(!e.considered) 
        SG.del(e); 
|\label{ln:triangle-once-2}|      else e.considered = true; 
|\vspace{0.5em}|} } }

|\vspace{0.25em}|/*********** |\textbf{\ul{Single-vertex compression kernel}}| (|\cref{sec:kernel-vertex}|) ***************/
|\label{ln:single-1}|low_degree(V v) {
  if(v.deg==0 or v.deg==1) 
|\label{ln:single-2}|    atomic SG.del(v); 
|\vspace{0.5em}|}

|\vspace{0.25em}|/************* |\textbf{\ul{Subgraph compression kernels}}| (|\cref{sec:sg-sp}|) ****************/
|\label{ln:spanners-1}|derive_spanner(vector<V> subgraph) { //Details in |\cref{sec:sg-spanners}|
  //Replace "subgraph" with a spanning tree
  subgraph = derive_spanning_tree(subgraph);
  //Leave only one edge going to any other subgraph.
  vector<set<V>> subgraphs(SG.sgr_cnt);
  foreach(E e: SG.out_edges(subgraph)) {
    if(!subgraphs[e.v.elem_ID].empty()) 
      atomic del(e); 
|\vspace{0.5em}||\label{ln:spanners-2}|} } 

|\label{ln:summaries-1}|derive_summary(vector<V> cluster) { //Details in |\cref{sec:sg-summaries}|
  //Create a supervertex "sv" out of a current cluster:
  V sv = SG.min_id(cluster); 
  SG.summary.insert(sv); //Insert sv into a summary graph
  //Select edges (to preserve) within a current cluster:
  vector<E> intra = SG.summary_select(cluster, SG.$\epsilon$);
  SG.corrections_plus.append(intra);
  //Iterate over all clusters connected to "cluster":
  foreach(vector<V> cl: SG.out_clusters(out_edges(cluster))) {
    [E, vector<E>] (se, inter) = SG.superedge(cluster,cl,SG.$\epsilon$);
    SG.summary.insert(se);
    SG.corrections_minus.append(inter); 
  }
|\label{ln:summaries-2}|  SG.update_convergence(); 
} 
\end{lstlisting}

\iftr
|\label{ln:spanners-1}|derive_spanner(vector<Vertex> subgraph) { //Details in |\cref{sec:sg-spanners}|
  //Replace "subgraph" with a spanning tree
  subgraph = derive_spanning_tree(subgraph);
  //Leave only one edge going to any other subgraph.
  vector<set<Vertex>> subgraphs(SG.sgr_cnt);
  foreach(Edge e: SG.out_edges(subgraph))
|\label{ln:spanners-2}|    if(!subgraphs[e.v.elem_ID].empty()) atomic del(e); } } }

|\add|  vector<bool> clusters(SG.sgr_cnt);
|\add|  foreach(Edge e: SG.out_edges(cluster)) {
|\add|    if(!clusters[e.v.elem_ID]) atomic {
|\add|      if(SG.if_create_superedge(e.u.elem_ID, e.v.elem_ID)) atomic del(e);
|\add|      Edge superedge = create_superedge(e.u.elem_ID, e.v.elem_ID);
|\add|      supervertex.append(superedge);
|\add|      clusters[e.v.elem_ID] = true;
\fi


\subsection{Single-Edge Kernels}
\label{sec:sg-ss}

We start from a simple kernel where the Slim Graph programming model provides the
developer with access to each edge together with the adjacent vertices and
their properties, such as degrees. In Slim Graph, we use this kernel to
express two important classes of compression schemes: spectral sparsification
and random uniform sampling.


\subsubsection{\macb{\underline{Spectral Sparsification with Slim Graph}}}
\label{sec:kernel-spectral}

\ 
%
%
%
%
In spectral
sparsification~\cite{spielman2011spectral}, one removes a certain number of
edges from an input graph~$G$ so that the \emph{Laplacian quadratic form of the
resulting subgraph~$H$ is approximately the same as that of $G$}. This means
that spectral sparsification preserves the \emph{graph spectrum} (i.e., the
eigenvalues of the Laplacian~$L_G$ of a graph~$G$). $L_G$, an established
concept in graph theory, is defined as $L \equiv D - A$, where $A$ is $G$'s
adjacency matrix ($A(i,j) = 1$ iff there is an edge from a vertex~$i$ to a
vertex~$j$) and $D$ is $G$'s degree matrix ($D(i,i) = d_i$ and $D(i,j) = 0$ for
$i \neq j$).
Formally, $H$ is a
\textbf{$\sigma$--spectral sparsifier} of $G$ if

\begin{gather}
\frac{1}{\sigma} x^T L_H x \le x^T L_G x \le \sigma x^T L_H x, \quad \forall_{x \in \mathbb{R}^n}
\end{gather}

\noindent
where $L_G$ and $L_H$ are Laplacians of $G$ and $H$ and
$x^T L x$ is the Laplacian quadratic form of $L$. 
\emph{The graph spectrum
determines various properties, for example \textbf{bipartiteness} or
\textbf{spectral clustering coefficient}}, which may be important for Slim
Graph users.  
%
%
%
%
Now, there exist many works on spectral sparsifiers~\cite{zhang2018towards,
anderson2014efficient, feng2016spectral, wang2017towards, koutis2016simple,
chu2018graph, kyng2018matrix, calandriello2018improved,
jambulapati2018efficient, soma2019spectral, spielman2011spectral,
lee2018constructing, kelner2013spectral, spielman2011graph,
batson2013spectral}. We \emph{exhaustively analyzed these
works} and we identified a method that needs only $O(m+n)$
storage and $O(m)$ time (others require
$\Omega(n^2)$ storage or have large hidden constants).
This method assumes
that the input graph has \emph{high conductance} (it is, intuitively,
``well-knit'').  This is true for many today's real-world
graphs~\cite{iyer2018bridging, yacsar2018fast}; we thus settle on this scheme.

Here, edges are sampled according to probabilities different for each edge.
These probabilities are selected in such a way that \emph{every vertex in the
compressed graph has edges attached to it w.h.p.}.  The fraction $\Upsilon$ of
remaining edges adjacent to each vertex can be proportional to $\log
(n)$~\cite{spielman2011spectral} ($\Upsilon = p \log (n)$) or to the average
vertex degree~\cite{iyer2018bridging} ($\Upsilon = p m/n$); $p$ is a user
parameter.  Then, each edge $(u,v)$ stays in the compressed graph with
probability $p_{u,v} = \min(1, {\Upsilon}/{\min(d_u, d_v)})$.
If the output graph must be weighted, then we set $ W(u,v) = 1 / p_{u,v}$.
Now, one can prove that a graph compressed according to the presented scheme
\emph{preserves spectrum well}~\cite{spielman2011spectral}.

%
\paragraph{Slim Graph Implementation}
In the corresponding kernel \texttt{spectral\_sparsify}
(Lines~\ref{ln:spectral-1}--\ref{ln:spectral-2}), each edge
\texttt{e} (provided as the kernel argument) is processed concurrently.
\texttt{edge\_stays} (the probability $p_{i,j}$ of sampling~\texttt{e}) is
derived based on $\Upsilon$ (a parameter maintained in \texttt{SG} and pre-initialied by the user) and degrees of vertices
\texttt{u} and \texttt{v} attached to \texttt{e}. Then, \texttt{e} is either
atomically deleted or
appropriately re-weighted.

\subsubsection{\macb{\underline{Uniform Sampling with Slim Graph}}}
\label{sec:sg-ssp}

We also express and implement random uniform sampling in Slim Graph. Here, each edge remains in the graph with a
probability~$p$. This simple scheme can be used to rapidly 
compress a graph while preserving accurately the \emph{\textbf{number of
triangles}}~\cite{tsourakakis2009doulion}.

\iftr
\maciej{fix} , as shown in a past analysis~\cite{tsourakakis2009doulion}.
\fi


\sloppy
\paragraph{Slim Graph Implementation}
The kernel for this scheme is shown in
Lines~\ref{ln:uniform-1}--\ref{ln:uniform-2}.  Its structure is
analogous to \texttt{spectral\_sparsify}.  The main difference is that the
sampling probability \texttt{edge\_stays} ($p$) is identical for each edge.


\subsection{Triangle Kernels for Triangle Reduction}
\label{sec:sg-tr}

The next class of compression kernels uses triangles (3-cycles) as the
``smallest unit of graph compression''.  Triangle kernels implement
\emph{Triangle Reduction} (TR): a class of compression schemes that generalizes
past work~\cite{kalavri2016shortest}.  In TR, a graph is compressed by removing
certain parts of a selected \emph{fraction of triangles}, sampled u.a.r.
({uniformly at random}).  Specific triangle parts to be removed are specified
by the developer.  Thus, we ``reduce'' triangles in a specified way. 
%

We focus on triangles because -- as we show
later (\cref{sec:theory},
\cref{sec:evaluation}) -- TR is \emph{versatile}: \emph{removing certain parts of triangles does not
significantly impact a \textbf{surprisingly large} number of graph properties}.
For example, removing an edge from each triangle does not increase the \emph{\textbf{number
of connected components}}, while removing the maximum-weight edge from each triangle does not
change the \emph{\textbf{weight of the minimum spanning tree}}.
Second, the
relatively low computational complexity of mining all the triangles
($O(m^{3/2})$ or $O(n d^2)$), combined with the existing bulk
of work on fast triangle listing~\cite{yacsar2018fast,
shun2015multicore, wolf2017fast, tsourakakisfast, green2014fast,
wang2016comparative, date2017collaborative, hu2018high, polak2016counting},
enables lossy compression of even the largest graphs available today.
Further, numerous approximate schemes find \emph{fractions of all
triangles in a graph much faster than $O(m^{3/2})$ or $O(n
d^2)$}~\cite{iyer2018asap, pavan2013counting, mcgregor2016better,
jowhari2005new, cormode2017second, bera2017towards, buriol2006counting,
seshadhri2013fast,
seshadhri2015simpler, jha2015counting, eden2017approximately, fox2018finding,
seshadhri2014wedge, jha2015space}, further reducing the cost of lossy
compression based on 
TR.
%


In the basic TR variant, we select $p T$ triangles from a graph u.a.r., $p \in
(0;1)$. In each selected triangle, we remove $x$ edges ($x \in \{1,2\}$),
chosen u.a.r..  We call this scheme \textbf{Triangle $p$-$x$-Reduction}, where
$p$ and $x$ are input parameters.
%
%

We advocate the versatility, extensibility, and flexibility of TR by discussing
variants of the basic TR scheme that enable tradeoffs between compression
performance, accuracy in preserving graph properties, and storage reductions. 
%
%
One variant is \textbf{\emph{Edge-Once} Triangle $p$-$x$-Reduction} (EO
$p$-$x$-TR).  Here, we consider each edge \emph{only once} for removal.  When a
triangle is selected for reduction for the first time (by some kernel
instance), if a random edge is not removed, it will \emph{not} be considered
for removal in another kernel instance.  This protects edges that are a part of
  \emph{many} triangles (that would otherwise be considered for deletion more
  often) and thus \emph{may be more important, e.g., they may be a part of
  multiple \textbf{shortest paths}}. 
Another example is EO $p$-1-Triangle Reduction with a modification in which we
remove an edge with the highest weight. \emph{This preserves the \textbf{exact
weight of the minimum spanning tree}}.


Certain Slim Graph users may be willing to sacrifice more accuracy in exchange
for further storage reductions.  In such cases, we offer \textbf{Triangle
  $p$-2-Reduction}.
Finally, we propose the \textbf{Triangle $p$-Reduction by Collapse} scheme in
which triangles \emph{are collapsed to single vertices}, each with a
probability $p$. This scheme changes the vertex set in addition to the edge
set, offering even more storage reduction.

\iftr
Our TR schemes come with additional parameters to control the scope of lossy
compression beyond parameter~$p$. For example, certain parameters
enable selecting whether or not an edge is removed \emph{\textbf{based on the
degree of adjacent vertices}}. 
For example, this enables 
the user to
preserve graph clusters 
or inter-cluster paths (by deleting fewer edges from dense and sparse graph
regions, respectively).
\fi

%
%


\paragraph{Slim Graph Implementation}

The kernel for the basic TR scheme (for $x=1$) is in
Lines~\ref{ln:triangle-1}--\ref{ln:triangle-2}; the EO variant is
presented in
Lines~\ref{ln:triangle-once-1}--\ref{ln:triangle-once-2}. In both
cases, the kernel argument \texttt{triangle} is implemented as a vector of
edges.  \texttt{SG}.$p$ is a probability of sampling a triangle.  We select an
edge to be removed with \texttt{rand} (an overloaded method that returns -- in
this case -- a random element of a container provided as the argument).  Here,
by selecting an edge for removal in a different way, one could
straightforwardly implement other TR variants.  For example, selecting an edge
with a maximum weight (instead of using \texttt{rand(triangle)}) would preserve
the MST weight.
The deletion is performed with the overloaded \texttt{SG.del} method.

\iftr
For additional control over the scope of compression,
we provide two parameters, $u$ and
$l$. Consider a triangle $X$ that is about to be ``reduced''. While any
vertex in $X$ has its degree bigger or equal than $u$, we remove both edges in
$X$. For triangles with vertex degrees between $u$ and $l$, we remove only one
edge (that belongs to the fewest triangles).  If $X$ has only vertices with
degrees lower than $l$, we stop sparsification.
\emph{This scheme enables users who have more detailed knowledge of the structure of
the graph to be compressed to tune $u$ and $l$ so that a desired fraction of
the edges is more or less sparsified.}
\fi

\iftr
Certain Slim Graph users may be willing to sacrifice more accuracy in exchange
for further storage reductions, beyond gains offered by the Triangle
  $p$-$x$-Reduction with $x \in \{2,3\}$. Here, we propose the \textbf{Triangle
  $p$-Reduction by Collapse} (TRC) scheme in which triangles \emph{are
  collapsed to single vertices}, each with a probability $p$. This scheme
  changes the vertex set in addition to the edge set and thus will require more
  sophisticated tools for analyzing the accuracy of compression.
\fi

\iftr
In \textbf{\emph{Min-Count} Triangle
$p$-$x$-Reduction} (MC Triangle $p$-$x$-Reduction), for each edge~$e$, we first
derive the number of triangles that $e$ belongs to.  This can be done
using simple triangle counting. Afterwards, we conduct the standard
{Triangle $p$-$x$-Reduction} step, but for each triangle to be reduced,
we remove the edge (or edges) which belongs to the \emph{fewest triangles}.
Intuitively, edges belonging to more triangles are more important and
it is more advantageous to prioritize the deletion of edges being in fewer
triangles.
\fi

\subsection{Single-Vertex Kernels}
\label{sec:kernel-vertex}


We enable the user to modify a single vertex. Our example kernel
(Lines~\ref{ln:single-1}-\ref{ln:single-2}) removes all vertices with degree
zero and one. The code is intuitive and similar to above-discussed edge kernels.
This enables compressing a graph while preserving the exact values of
\textbf{\emph{betweenness centrality}}, because degree-1 vertices do not
contribute any values to shortest paths between vertices with degrees higher
than one~\cite{sariyuce2013betweenness}.

\subsection{Subgraph Kernels}
\label{sec:sg-sp}


Slim Graph allows for \textbf{executing a kernel on an arbitrary
\emph{subgraph}}. \emph{This enables expressing and implementing different
sophisticated compression schemes}, such as spanners (graphs that preserve
\textbf{pairwise distances}) and lossy graph summarization (graphs
that preserve \textbf{neighborhoods}).



\subsubsection{\macb{\underline{Overview of Slim Graph Runtime}}}
\label{sec:sg-mappings}

To clarify subgraph kernels, we first
summarize the general Slim Graph runtime execution, see
Listing~\ref{lst:general}. After initializing \texttt{SG}, 
assuming subgraph kernels are used, Slim Graph constructs \texttt{SG.mapping}, a structure
that maps each vertex to its subgraph. Mappings are discussed
in~\cref{sec:sg-mappings}; they enable \emph{versatility and flexibility}
in implementing lossy compression schemes in Slim Graph.
Next, a function \texttt{run\_kernels} executes each kernel concurrently. These
two steps are repeated until a convergence condition is achieved. The convergence
condition (and thus executing all kernels more than once) is only necessary for
graph summarization. All other lossy compression schemes expressed in Slim
Graph require only a single execution of \texttt{run\_kernels}.


\begin{lstlisting}[aboveskip=0em,abovecaptionskip=-0.05em,belowskip=0em,float=h,label=lst:general,caption=\textmd{
Overview of Slim Graph runtime execution using an example of subgraph kernels.}]
SG.init($G$); //Init the SG object using the input graph $G$.

/* In addition, here the user can initialize various parameters
related to the selected lossy compression, etc. */
while(!SG.converged) { //"converged" is updated in "run_kernels"
  if(SG.kernel == SUBGRAPH) SG.construct_mapping();
  SG.run_kernels(); //Execute all kernels concurrently
}

SG.free(); //Perform any necessary cleanup.
\end{lstlisting}

\subsubsection{\macb{\underline{Mappings}}}
\label{sec:sg-mappings}

\iftr
-- during our exhaustive analysis
of existing lossy compression methods -- 
\fi


%
%
%
While analyzing lossy graph compression, we discovered that
many representative spanner and graph summarization schemes first decompose a
graph into disjoint subgraphs. Next, these schemes use the obtained intra- and
inter-subgraph edges to achieve higher compression ratios or to ensure that the
compression preserves some graph properties (e.g., diameter).  Details of such
graph decompositions are algorithm-specific, but they can \emph{all} be defined
by a mapping that assigns every vertex to its subgraph.
Thus, to express any such compression algorithm in Slim Graph, we enable
constructing arbitrary mappings.


\paragraph{Example Mappings}
%
%
Two important mappings used in Slim Graph are based on low-diameter
decomposition~\cite{miller2015improved} (takes $O(n+m)$ work) and clustering
based on Jaccard similarity~\cite{real1996probabilistic} (takes $O(mN)$ work;
$N$ is \#clusters). In the former (used for spanners), resulting
subgraphs have (provably) low diameters.  In the latter (used for graph
summarization), resulting subgraphs consist of vertices that are similar to one
another with respect to the Jaccard measure.  Both schemes are extensively
researched and we omit detailed specifications.

\iftr
An example straightforward mapping assigns a vertex to a
corresponding connected component. There are numerous
established schemes for deriving connected components~\cite{}
in $O(n+m)$ work.
\fi
%


\paragraph{Implementing Mappings}
%
%
To develop mappings, a user can use either the established vertex-centric
abstraction or simply access the input
graph (maintained as adjacency arrays) through the \texttt{SG} container.
Implementation details are straightforward;
they directly follow algorithmic specifications of low-diameter
decompositions~\cite{miller2015improved} or clustering~\cite{shin2019sweg}.
From populated mappings, the Slim Graph runtime derives subgraphs
that are processed by kernel instances.

\iftr
\begin{lstlisting}[aboveskip=1em,float=!h,label=lst:mappings,caption=\textmd{
Slim Graph execution for subgraph kernels that enable expressing
and implementing spanners and graph summarization. $G$ is the input graph.}]
void SG.construct_mapping() {
  SG.mapping = new vector<pair<Vertex,int>>();
}

mapping(vector<Vertex> vertices, vector<Edge> edges, vector<vector<Vertex>> subgraphs) {
  foreach(Edge e: edges) {
    
  }
}
\end{lstlisting}
\fi

\iftr
\subsubsection{\macb{\underline{\hll{Iterative Execution}}}}
\label{sec:sg-iterative}
\fi

\subsubsection{\macb{\underline{Spanners with Slim Graph}}}
\label{sec:sg-spanners}


An \textbf{$(\alpha,\beta)$-spanner}~\cite{peleg1989optimal} is a subgraph $H =
(V,E')$ of $G=(V,E)$ such that $E' \subset E$ and

$$
dist_G(u,v) \le dist_H(u,v) \le \alpha \cdot dist_G(u,v) + \beta,\quad \forall u,v \in V.
$$

We exhaustively analyzed works on
spanners~\cite{elkin2018efficient, lenzen2018centralized, censor2018sparsest,
censor2018distributed, dory2018distributed, parter2018local,
alstrup2017constructing, ahn2012graph, pettie2010distributed,
baswana2010additive, althofer1993sparse, baswana2007simple, peleg1989graph,
miller2015improved} 
and we select a state-of-the-art scheme by Miller et al.~\cite{miller2015improved}
that provides best known work-depth bounds and is easily parallelizable. It first decomposes 
a graph into low-diameter subgraphs.
An input parameter $k \ge 1$ controls how
large these subgraphs are.
Then, 
it derives a spanning tree of each 
subgraph; these trees have low diameters (${k}\log (n)$ w.h.p.).
Thus, each
such subgraph is sparsified and
$k$ controls the scope of this sparsification.
Formally, for an input parameter $k \ge 1$, one obtains an $O(k)$-spanner with
$O(n^{1+1/k})$ edges.  For this, $G$ is partitioned into clusters of the
diameter at most ${k}\log n$ with probability at least $1 - 1/n^{k-1}$. The
computation takes $O(\log n \log^* n)$ depth and $O(m)$ work in the PRAM model.

After that, for each subgraph~$C$ and each vertex~$v$ belonging to $C$, if $v$
is connected to any other subgraph with edges $e_1, ..., e_l$, only one of these
edges is added to the resulting $O(k)$-spanner
that has $O(n^{1+1/k})$ edges.



\paragraph{Slim Graph Implementation}
The corresponding kernel is in
Lines~\ref{ln:spanners-1}--\ref{ln:spanners-2}
  (Listing~\ref{lst:kernels}).
First, one derives a spanning tree of \texttt{subgraph} that is the argument of
the compression kernel \texttt{derive\_spanner}.  Then, by iterating over
edges outgoing from \texttt{subgraph}, the implementation leaves only one edge
between any two subgraphs (here, we use \texttt{sgr\_cnt}, a field of
\texttt{SG} that maintains the number of subgraphs.

\iftr
We exhaustively analyzed works on
spanners~\cite{elkin2018efficient, lenzen2018centralized, censor2018sparsest,
censor2018distributed, dory2018distributed, parter2018local,
alstrup2017constructing, ahn2012graph, pettie2010distributed,
baswana2010additive, althofer1993sparse, baswana2007simple, peleg1989graph,
miller2015improved} and we identified schemes that are easily parallelized~\cite{elkin2018efficient, miller2015improved}. 
Here, the graph is first decomposed into subgraphs. Then, each
  subgraph is sparsified according to a certain algorithm. Finally, selected
  edges between subgraphs are also removed. An input parameter $k \ge 1$ controls how
  large these subgraphs are. 
\fi

\iftr
To ensure good spanning properties, these trees have low diameters (${k}\log (n)$ w.h.p.).
After that, for each cluster~$C$ and each vertex~$v$ belonging to $C$, if $v$
is connected to any other cluster with edges $e_1, ..., e_l$, only one of these
edges is added to the resulting spanner.
This ultimately gives an $O(k)$-spanner with
$O(n^{1+1/k})$ edges.
\fi

\iftr
 Formally, for an input parameter $k \ge 1$, one obtains an $O(k)$-spanner with
$O(n^{1+1/k})$ edges.  For this, $G$ is partitioned into clusters of the
diameter at most ${k}\log n$ with probability at least $1 - 1/n^{k-1}$. The
computation takes $O(\log n \log^* n)$ depth and $O(m)$ work in the PRAM model.
After that, for each cluster~$C$ and each vertex~$v$ belonging to $C$, if $v$
is connected to any other cluster with edges $e_1, ..., e_l$, only one of these
edges is added to the resulting spanner.
\fi

\subsubsection{\macb{\underline{Lossy Summaries with Slim Graph}}}
\label{sec:sg-summaries}

\sloppy
We enable Slim Graph to support \textbf{lossy
$\epsilon$-summarization ($\epsilon$-summaries)}.  The general idea behind
these schemes is to \emph{summarize} a graph by merging specified subsets of
vertices into \emph{supervertices}, and merge parallel edges between
supervertices into \emph{superedges}.  A parameter $\epsilon$ bounds the error
(details are algorithm-specific).
We exhaustively analyzed existing 
schemes \cite{riondato2017graph,
liu2018graph, shin2019sweg, campinas2013efficiency, toivonen2011compression,
chen2009mining, liu2014distributed, fan2012query,
toivonen2012network, sadri2017shrink,
zhou2017summarisation, navlakha2008graph, tsalouchidou2018scalable,
beg2018scalable, kumar2018utility, dunne2013motif}. We focus on SWeG, a
recent scheme~\cite{shin2019sweg} that constructs supervertices with a
generalized Jaccard similarity.


\paragraph{Slim Graph Implementation}
The corresponding kernel is in
Lines~\ref{ln:summaries-1}--\ref{ln:summaries-2}
  (Listing~\ref{lst:kernels}).  It first creates a supervertex
  \texttt{sv} out of a processed cluster; \texttt{sv} is added to the
  summary graph.  Next, an algorithm-specific
  \texttt{summary\_select} method returns edges selected from \texttt{cluster};
  $\epsilon$ determines the scope of lossy compression (i.e., how many
  intra-cluster edges are irreversibly dropped). The returned edges are kept
  in a data structure \texttt{corrections\_plus}
  (they are used to better preserve neighborhoods). Finally, 
  one iterates over neighboring clusters (using simple predefined
  methods that appropriately aggregate edges).  For each neighboring cluster, a
  superedge may be created inside method \texttt{SG.superedge}.  This method
  (1) drops certain sampled inter-cluster edges (for lossy compression), (2)
  returns a newly-created superedge \texttt{se} (or a null object, if no
  superedge was created), and (3) a vector \texttt{inter} with edges
  that \emph{do not belong to the created superedge} (assuming \texttt{se} is
  created) and thus \emph{must be removed whenever one accesses edges that form superedge~\texttt{se}}.
Thus, edges in \texttt{inter} are added to \texttt{corrections\_minus}, a data structure
  with corrections.

\subsection{Slim Graph vs Other Schemes}
\label{sec:sg-ot}

Other forms of lossy graph compression
could be used in future Slim Graph versions
as new compression kernels.
%
%
First, \textbf{cut sparsifiers}~\cite{benczur1996approximating} only target the problem of graph cuts
Formally, in cut sparsification~\cite{benczur1996approximating}, one approximates an 
input graph~$G$ with a graph~$H$ such that, for every subset of vertices $U
\subset V$, the weight of the edges leaving $U$ is approximately the same in
$G$ as in the sparsifier $H$. 
{We do not use cut sparsifiers as they are NP-Hard to derive and
because they are a special case of spectral sparsifiers:}
a given $G$ and its cut sparsifier $H$ must satisfy the same set
if inequalities as in spectral sparsification, but only for $x \in \{0,1\}^n$.
Second, other schemes target specifically \emph{dynamic and weighted graphs}~\cite{liu2012compressing, henecka2015lossy}.
Third, \textbf{low-rank approximation}~\cite{savas2011clustered} of clustered
Singular Value Decomposition (SVD) was shown to yield very high error
rates~\cite{sui2012parallel, savas2011clustered}; we confirm this
(\cref{sec:evaluation}).  Moreover, it has a prohibitive time and space
complexity of $O\left(n_c^3\right)$ and $O(n_{c}^2)$ where $n_c$ is the size of
the largest cluster $n_c \in O(n)$.
Finally, \textbf{lossy summarization} based on the \textbf{Minimum Description
Length principle}~\cite{navlakha2008graph} and \textbf{Lossy Linearization}~\cite{maserrat2012community} have
high time complexities of $O(m^2 \log n)$ and $O(m d I T)$, respectively, making them infeasible for
today's graphs.
These schemes could be used
in future Slim Graph versions.
%
%



\subsection{Kernel Strengths: Takeaways}


Compression kernels are \emph{simple}: the ``local'' (e.g., vertex-centric)
view of the graph simplifies designing compression algorithms.
Slim Graph implementations of compression schemes based on vertex, edge, or
triangle kernels use 3--10$\times$ fewer lines of code than the 
corresponding standard baselines. Subgraph
kernels use up to 5$\times$ fewer code lines (smaller gains
are due to the fact that compression schemes that must be expressed with subgraph kernels
are inherently complex and some part of this complexity must also be implemented within
Slim Graph mappings).
Second, kernels
are \emph{flexible}: one easily extends a kernel to cover
a different graph property (e.g., preserving the exact MST weight with TR only
needs removing an edge with the highest weight).
%
%
Third, different kernels offer a \emph{tradeoff in compression speed,
simplicity, and flexibility}. Vertex kernels have limited expressiveness
(as is vertex-centric graph processing~\cite{salihoglu2014optimizing, yan2014pregel}), but they are simple to
use and reason about, and running all vertex kernels takes $\Omega(n)$ work.
Edge kernels are less limited but they take $\Omega(m)$ work. Triangle kernels
are even more expressive but take $O(m^{3/2})$ work.
\iftr
On the other hand, Slim Graph does \emph{not} iterate over all possible
subgraphs. Instead, the user provides a mapping that determines the
partitioning of the graph into subgraphs.  
\fi
Finally, 
as mappings are arbitrary,
subgraph kernels are the most expressive but also complex to use. We recommend
using them if global knowledge of the
graph structure is needed. Currently, we use them with 
spanners and summarization. 
Another possible use case left for
future work are spectral sparsifiers that do not assume graph's high
conductance.


\section{Slim Graph: Accuracy Metrics}
\label{sec:metrics}


We now establish metrics for assessing the impact of graph compression on 
algorithm outcomes. 
we present different metrics that can be used with algorithms that
provide different types of output (e.g., a scalar number or a vector, cf.~the
``output type'' column in~Table~\ref{tab:problems}).
%
%
\iftr
\maciej{fix}
We present the most interesting metrics. However, in Slim
Graph, we also use simple tools such as relative changes in the scalar output
of graph algorithms (e.g., the change in the number of triangles).
\fi
\emph{Our metrics are {generic} and can be used with any compression methods.}
%
%
%



\subsection{Algorithms with Scalar Output}
\label{sec:metrics_scalar}



The first and simplest accuracy metric, $\mathcal{A}_{S} =
|{\widetilde{R}} / {R}|$, measures the relative change of a \emph{S}calar
output of graph algorithms
(output type ``S'' in Table~\ref{tab:problems}), for example the total number
of triangles.
%
%
%
$\widetilde{R}$ and $R$ are the outcome for the approximated and the original
graph, respectively. For example, $\widetilde{R}$ and $R$ can be the number of
triangles in the approximated and original graph, respectively.
Table~\ref{tab:metrics} presents example interpretations of $\widetilde{R}$ and
$R$ in selected graph algorithms.

\begin{table}[h]
\centering
\footnotesize
\sf
\begin{tabular}{ll@{}}
\toprule
%
%
\textbf{Algorithm} & \textbf{Interpretation of $\widetilde{R}$ and $R$} \\
\midrule
Triangle Counting & Number of triangles \\
Connected Components & Number of connected components \\
Minimum Spanning Tree & Weight of the minimum spanning tree\\
Maximum Matching & Cardinality of the maximum matching~\cite{besta2019substream}\\
Minimum Edge Cover & Cardinality of the minimum edge cover\\
%
%
Graph Coloring & \makecell[l]{The number of colors used to color the graph} \\
Minimum Vertex Cover & Cardinality of the minimum vertex cover \\
Maximum Independent Set & Cardinality of the maximum independent set \\
%
\bottomrule
\end{tabular}
\caption{\textmd{\textbf{The interpretation of $\widetilde{R}$ and $R$ for
selected graph algorithms.} Both $\widetilde{R}$ and $R$ always refer to the
same algorithm result (e.g., the number of triangles); the difference is that
the former refers to this property in the approximated graph while the latter
is associated with the original dataset.}}
%
\label{tab:metrics}
\end{table}

\subsection{Algorithms with Vector Output}
\label{sec:metrics_vector}

Next, we consider algorithms that output a vector where each element is a value
associated with one vertex (e.g., in PageRank); see the output type ``V'' in Table~\ref{tab:problems}.
These result vectors come with various semantics,
requiring different accuracy metrics.  
Thus, we discuss simple counts of reordered element pairs
(\cref{sec:metrics-v-reordered}), and more complex measures such as statistical
divergences (\cref{sec:metrics-v-divergences}) and algorithm-specific measures
(\cref{sec:metrics-v-specific}).

\subsubsection{Counts of Reordered Pairs}
\label{sec:metrics-v-reordered}

We first consider algorithms that output a vector where each element is a value
associated with one vertex (e.g., in PageRank).
The $\mathcal{A}_{RE}$ metric tracks the number of vertex pairs
that are \emph{RE}ordered with respect to the considered score such as rank. We
have
%

$$
\mathcal{A}_{RE} = |P_{RE} / {\binom{n}{2}}|
$$

where $P_{RE}$ is the number of vertex pairs that are reordered after applying
compression; we divide it by the maximum possible number of reordered pairs
$\binom{n}{2}$.
%

\iftr
\vspace{-1em}
\begin{gather}
\mathcal{A}_{RE} = \left|1 - P_{RE} / {\footnotesize\binom{n}{2}}\right|
\end{gather}
\fi

\subsubsection{Counts of Reordered Neighbors}
\label{sec:metrics-v-reordered-n}

In addition to $\mathcal{A}_{RE}$, we also use a metric $\mathcal{A}_{REN}$
that counts \emph{RE}ordered \emph{N}eighboring vertices. This metric is much
more efficient to compute, taking $O(m)$ time instead of $O\left(n^2\right)$.

\subsubsection{Statistical Divergences}
\label{sec:metrics-v-divergences}

Some graph properties and results of algorithms can be modeled with certain
\emph{probability distributions};
see the output type ``P'' in
Table~\ref{tab:problems}.
For example, in PageRank, one assigns each vertex (that models a web page) the
probability (rank) of a random surfer landing on that page.
Another example is degree distribution: the probability of each vertex
having a certain degree.
In such cases, \emph{we observe that one can use the concept of a
\textbf{divergence}: a statistical tool that measures the distance between
probability distributions}.
Divergence generalizes the notion of ``distance'': it does not need not be
symmetric and need not satisfy the triangle inequality.  There are dozens of
divergences~\cite{cha2007comprehensive, basseville2010divergence}; many
belong to two groups: so called $f$-divergences and Bregman
divergences~\cite{basseville2010divergence}.

In order to develop Slim Graph, we analyzed various divergences to understand
which one is best suited for Slim Graph.  We select the \emph{Kullback-Leibler (KL)
divergence}~\cite{kldivergence}, which originated in the field of information
theory.  The reasons are as follows.  First, the Kullback-Leibler divergence is
generic and applicable to many problems as it is \emph{the only Bregman
divergence which is also an $f$-divergence}~\cite{kldivergence}.  Moreover, it
has been used to measure the information loss while approximating probability
distributions~\cite{cover2012elements, kldivergence}.  Finally, it
has recently been used to find differences between brain networks by analyzing
distributions of the corresponding graph
spectra~\cite{takahashi2012discriminating}. \emph{Thus,
Kullback-Leibler divergence can be used to analyze the information loss in
graphs compressed by Slim Graph when considering graph properties such as
PageRank distributions.}

Formally, Kullback-Leibler divergence measures the deviation of one
probability distribution from another one. The deviation of distribution $Q$ from
$P$ is defined as $\Sigma_i P(i)\log_2\frac{P(i)}{Q(i)}$.

\iftr
\vspace{-1em}
\begin{gather}
D_{KL}(P || Q)=-\sum_i P(i)\log_2\frac{Q(i)}{P(i)}
\end{gather}
\fi
%
%
%
The Kullback-Leibler divergence is a non-negative
number, equal to zero if and only if $P$ and $Q$ are identical. 
%
%
The lower Kullback-Leibler divergence between probability
distributions is, the closer a compressed graph is to the original one,
regarding the considered probability distribution.

\iftr


\setlength{\tabcolsep}{3pt}
\renewcommand{\arraystretch}{0.7}
\centering
 \ssmall
\sf
\begin{tabular}{lrrrrrrrrrrrr}
\toprule
 & $|V|$ & $|E|$ & \makecell[c]{Shortest\\$s$-$t$ path length} & \makecell[c]{Average\\path length} & Diameter & \makecell[c]{Average\\degree} & \makecell[c]{Maximum\\degree}  & \#Triangles & \makecell[c]{\#Connected\\components} & \makecell[c]{Chromatic\\number} & \makecell[c]{Max. indep.\\set size} & \makecell[c]{Max. cardinal.\\matching size}\\
\midrule
\makecell[l]{Original graph} & $n$ & $m$ & $\mathcal{P}$ & $\overline{P}$ & $D$ & $\overline{d}$ & $d$ & $T$ & $\mathcal{C}$ & $C_R$ & $\widehat{I}_S$ & $\widehat{M}_C$ \\ 
\midrule

\rowcolor{llyellow} \makecell[l]{Lossy $\epsilon$-Summary} & $n$ & $ m \pm 2\epsilon m$ & $1, \dotsc, \infty$ & $1, \dotsc, \infty$ & $1, \dotsc, \infty$ & $\overline{d} \pm \epsilon \overline{d} $ & $d \pm \epsilon d $ & $T \pm 2\epsilon m$ & $ \mathcal{C} \pm 2\epsilon m$ & $ C_R \pm 2\epsilon m$ & $\widehat{I}_S \pm 2\epsilon m$ & $\widehat{M}_C \pm 2\epsilon m$ \\

\makecell[l]{Simple $p$--sampling} & $n$ & $(1-p) m$ & $\infty$ & $\infty$ & $\infty$ & $(1-p)\overline{d}$ & $(1-p)d$ & $ \left(1 - p^3\right)T$ & $\le \mathcal{C} + pm$ & $\geq C_R - pm$ & $\leq \widehat{I}_S + pm$ & $\geq \widehat{M}_C - pm$ \\

\makecell[l]{Spectral $\epsilon$ sparsifier} & $n$ & $\tilde O\left(\frac{n}{\epsilon^2}\right)$ &  $\leq n$ & $\leq n$ & $\leq n$ & $\tilde O\left(\frac{1}{\epsilon^2}\right)$ & $\geq \frac{d}{2(1+\epsilon)}$ & $\tilde O\left(\frac{n^{3/2}}{\epsilon^3}\right)$ & $\overset{w.h.p.}{=} \mathcal{C}$ & $\leq \frac{d}{2(1+\epsilon)}$ & $\geq \frac{2(1+\epsilon)n}{d}$   & $\geq 0 $\\

$O(k)$--spanner & $n$ & $O(n^{1+1/k})$ & $O( k \mathcal{P})$ & $O(k \overline{P})$ & $O( k D)$ & $O(n^{1/k})$ & $\leq d$ & $O(n^{1+2/k})$ & $\mathcal{C}$ & $O(n^{1/k}\log n)$ & $\Omega\left(\frac{n^{1-1/k}}{\log n} \right)$ &  $\geq 0 $ \\

\makecell[l]{EO $p$--1--Triangle Red.} & $n$ & $\le m - \frac{pT}{ 3d }$ & $\overset{w.h.p.}{\le} \mathcal{P} + p \mathcal{P}$ & $\le \overline{P} + \frac{p T}{n(n-1)}$ & $\overset{w.h.p.}{\le}  D + p D$ & $\le \overline{d} - \frac{pT}{ d n}$  & $\geq d/2$ &  $ \le (1-\frac{p}{d}) T$ & $\mathcal{C}$ & $\ge C_R - p T$ & $\le \widehat{I}_S + p T$ & $\geq \widehat{M}_C / 2$ \\

\makecell[l]{remove $k$ deg-$1$ vertices} & $n-k$ & $m-k$ & $\mathcal{P}$ & $\geq \overline{P} - \frac{kD}{n} $ & $\geq D-2$ & $\geq \overline{d}-\frac{k}{n}$  & $d$ &  $T$ & $\mathcal{C}$ & $C_R$ & $\geq \widehat{I}_S - k$ & $\geq \widehat{M}_C -k$ \\

\bottomrule
\end{tabular}
%
\caption{
\textmd{\textbf{The impact of various compression schemes on the outcome of
selected graph algorithms.} Bounds that do not include inequalities hold
deterministically. If not otherwise stated, the other bounds hold in
expectation. Bounds annotated with w.h.p. hold w.h.p. (if the
involved quantities are large enough). Note that since the listed compression schemes (except the scheme where we remove the degree $1$ vertices and $\epsilon$-summaries) return a subgraph of the original graph, $m$, $C_R$, $\overline{d}$, $d$, $T$, and $\widehat{M}_C$ never increase. Moreover, $\mathcal{P}$, $\overline{P}$,$D$, $\mathcal{C}$, and $\widehat{I}_S$ never decrease during compression. $\epsilon$ is a parameter that controls how well a spectral sparsifier approximates the original graph spectrum.}
}
\vspace{-3.5em}
\label{tab:theory-table}
\end{table*}

\fi

\subsection{Algorithm-Specific Measures: BFS}
\label{sec:metrics-v-specific}

%
BFS is of particular importance in the HPC community as it is commonly used to
test the performance of high-performance systems for irregular workloads, for
example in the Graph500 benchmark~\cite{murphy2010introducing}.
Much focus was placed on developing fast and scalable BFS
routines~\cite{fu2014parallel, ueno2013parallel, merrill2012scalable,
luo2010effective, zou2013direction, bisson2016parallel, checconi2014traversing,
yasui2015fast, beamer2013direction, buluc2017distributed,
beamer2013distributed, schardl2010design, berrendorf2014level,
leiserson2010work, yasui2013numa, yoo2005scalable, bulucc2011parallel,
xia2009topologically, gazit1988improved, hong2011efficient,
Satish:2012:LEG:2388996.2389015}.
BFS is also a special case for Slim Graph metrics.  Its
outcome that is important for Graph500 is a vector of \emph{predecessors} of
every vertex in the BFS traversal tree. Thus, we cannot use simple metrics for
vector output 
as they 
are suitable for centrality-related graph problems where a swapped pair
of vertices indicates that a given compression scheme impacts vertex ordering;
no such meaning exists in the context of vertex predecessors.  Moreover, we
cannot use divergences because a vector of predecessors does not form a
distribution.

\begin{figure}[h]
\centering
\includegraphics[width=0.48\textwidth]{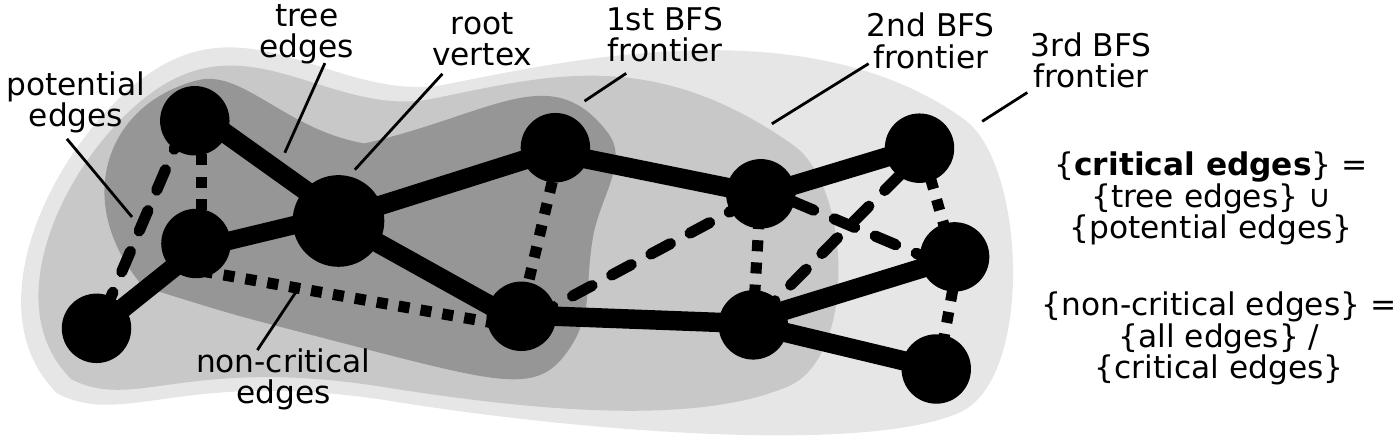}
\vspace{-2em}
\caption{\textmd{Edge types considered in Slim Graph
when analyzing the outcome of BFS.
Tree edges constitute a given BFS traversal tree. Potential edges are the edges
that may have been included in the BFS tree. These both types constitute critical edges.
}}
\label{fig:lossy}
%
\end{figure}

To understand how a given compression scheme impacts the BFS outcome, we
first identify various types of edges used in BFS. The core
idea is to identify how many \emph{critical} edges that may constitute the BFS
tree are preserved after sparsification.
For a given BFS traversal, the set of critical edges
$E_{cr}$ contains the edges from the actual output BFS traversal tree
(\emph{tree edges}) \emph{and} the edges that could potentially be included in
the tree by replacing any of the tree edges (\emph{potential edges}). We
illustrate an example in Figure~\ref{fig:lossy}.
$\widetilde{E}_{cr}$ are critical edges in the compressed graph, for a traversal starting
from the same root.
Now, the fraction $|\widetilde{E}_{cr}| / |E_{cr}|$ indicates the change in the number
of critical edges.

%
%
%
%
%
%

\begin{landscape}
\begin{table}
%
\setlength{\tabcolsep}{2.5pt}
\renewcommand{\arraystretch}{1.3}
\centering
\small
\sf
\begin{tabular}{lrrrrrrrrrrrr}
\toprule
 & $|V|$ & $|E|$ & \makecell[c]{Shortest\\$s$-$t$ path length} & \makecell[c]{Average\\path length} & Diameter & \makecell[c]{Average\\degree} & \makecell[c]{Maximum\\degree}  & \#Triangles & \makecell[c]{\#Connected\\components} & \makecell[c]{Coloring\\number} & \makecell[c]{Max. indep.\\set size} & \makecell[c]{Max. cardinal.\\matching size}\\
\midrule
\makecell[l]{Original graph} & $n$ & $m$ & $\mathcal{P}$ & $\overline{P}$ & $D$ & $\overline{d}$ & $d$ & $T$ & $\mathcal{C}$ & $C_G$ & $\widehat{I}_S$ & $\widehat{M}_C$ \\ 
\midrule

\vspace{0.25em}\makecell[l]{Lossy $\epsilon$-summary} & $n$ & $ m \pm 2\epsilon m$ & $1, \dotsc, \infty$ & $1, \dotsc, \infty$ & $1, \dotsc, \infty$ & $\overline{d} \pm \epsilon \overline{d} $ & $d \pm \epsilon d $ & $T \pm 2\epsilon m$ & $ \mathcal{C} \pm 2\epsilon m$ & $ C_G \pm 2\epsilon m$ & $\widehat{I}_S \pm 2\epsilon m$ & $\widehat{M}_C \pm 2\epsilon m$ \\

\makecell[l]{Simple $p$--sampling} & $n$ & $(1-p) m$ & $\infty$ & $\infty$ & $\infty$ & $(1-p)\overline{d}$ & $(1-p)d$ & $ (1 - p^3)T$ & $\le \mathcal{C} + pm$ & $\geq \frac{1-p}{2} C_G $ & $\leq \widehat{I}_S + pm$ & $\geq (1-p) \widehat{M}_C $ \\

\vspace{0.25em}\makecell[l]{Spectral $\epsilon$-sparsifier} & $n$ & $\tilde O({n}/{\epsilon^2})$ &  $\leq n$ & $\leq n$ & $\leq n$ & $\tilde O({1}/{\epsilon^2})$ & $\geq {d}/{2(1+\epsilon)}$ & $\tilde O({n^{3/2}}/{\epsilon^3})$ & $\overset{w.h.p.}{=} \mathcal{C}$ & $\geq 0$ & $\leq n$   & $\geq 0 $\\

$O(k)$--spanner & $n$ & $O(n^{1+1/k})$ & $O( k \mathcal{P})$ & $O(k \overline{P})$ & $O( k D)$ & $O(n^{1/k})$ & $\leq d$ & $O(n^{1+2/k})$ & $\mathcal{C}$ & $O(n^{1/k}\log n)$ & $\Omega\left(\frac{n^{1-1/k}}{\log n} \right)$ &  $\geq 0 $ \\

\vspace{0.25em}\makecell[l]{EO $p$--1--Triangle Red.} & $n$ & $\le m - \frac{pT}{ 3d }$ & $\overset{w.h.p.}{\le} \mathcal{P} + p \mathcal{P}$ & $\le \overline{P} + \frac{p T}{n(n-1)}$ & $\overset{w.h.p.}{\le}  D + p D$ & $\le \overline{d} - \frac{pT}{ d n}$  & $\geq d/2$ &  $ \le (1-\frac{p}{d}) T$ & $\mathcal{C}$ & $\ge \frac{1}{3} C_G $ & $\le \widehat{I}_S + p T$ & $\geq \frac{2}{3} \widehat{M}_C $ \\




\makecell[l]{remove $k$ deg-$1$ vertices} & $n-k$ & $m-k$ & $\mathcal{P}$ & $\geq \overline{P} - \frac{kD}{n} $ & $\geq D-2$ & $\geq \overline{d}-\frac{k}{n}$  & $d$ &  $T$ & $\mathcal{C}$ & $\geq C_G-1$ & $\geq \widehat{I}_S - k$ & $\geq \widehat{M}_C -k$ \\

\bottomrule
\end{tabular}
%
\caption{
\textmd{\textbf{The impact of various compression schemes on the outcome of
selected graph algorithms.} Bounds that do not include inequalities hold
deterministically. If not otherwise stated, the other bounds hold in
expectation. Bounds annotated with w.h.p. hold w.h.p. (if the
involved quantities are large enough). Note that since the listed compression schemes (except the scheme where we remove the degree $1$ vertices and $\epsilon$-summaries) return a subgraph of the original graph, $m$, $C_G$, $\overline{d}$, $d$, $T$, and $\widehat{M}_C$ never increase. Moreover, $\mathcal{P}$, $\overline{P}$,$D$, $\mathcal{C}$, and $\widehat{I}_S$ never decrease during compression. $\epsilon$ is a parameter that controls how well a spectral sparsifier approximates the original graph spectrum.}
}
%
%
\label{tab:theory-table}
\end{table}
\end{landscape}

\section{Theoretical Analysis}
\label{sec:theory}

We analyze theoretically how Slim Graph impacts
graph properties. 
Our main result are \emph{novel bounds} (more than 20 non-trivial ones)
for \emph{each combination of 12 graph properties and 7 compression schemes}. 
\iftr
\fi
%
%
%
%
%
We discuss the most interesting results;
Table~\ref{tab:theory-table}
summarizes our bounds. Next, we give an overview of the main insights,
comparing the various compression schemes with respect to the quantities they
manage to preserve and to how sparse the compressed graph is.

\subsection{Arbitrary Edge Sparsification}

We begin with bounds that hold for any compression scheme that removes edges
and returns a subgraph.

\subsubsection{Chromatic Number}

First, we can obtain \emph{lower bounds} on the chromatic number by counting
how many edges are deleted.
Specifically, deleting an edge decreases the chromatic number of a graph by at
most $1$. 
%
%
Consider a coloring of the graph after some edge $(u,v)$ is deleted. We can
assign a new color~$X$ to the vertex~$v$, identical to the color of~$u$, if no
other neighbor of~$v$ is already colored with~$X$. This potentially decreases
the total number of colors in the graph by $1$.
%

Moreover, when analyzing the \emph{upper bounds} on the chromatic number, we
can use the fact that every graph with maximum degree $d$ can be colored with
$d+1$ colors.  Now, whenever removing some edges decreases~$d$, this also
decreases the upper bound on the chromatic number by~$1$.

\subsubsection{Independent Set}

Deleting an edge increases the size of the largest independent set by at most
$1$. 

\subsection{Triangle Kernels: Edge-Once $p$-1-TR}

We now analyze Triangle Reduction.
The obtained bounds focus on Edge-Once $p$-1-Reduction, but they also hold for
slight variants of this scheme; the important thing is to ensure that no two
edges are deleted from the same triangle \emph{and} every edge is only
considered for deletion at most once. If edges can be considered for deletion
multiple times, the returned graph is still a $2$-spanner.

\subsubsection{Edge Count} 

We expect to sample $p T$ triangles. Now, each triangle shares an edge with at most
$3 d$ other triangles. To see this, consider any edge~$(v,u)$ of an arbitrary triangle.
If both $v$ and $u$ have each degrees lower than $d$, then clearly $(v,u)$ belongs to
fewer than $d$ triangles. If any of $v$ or $u$ has a maximum degree~$d$, then
similarly the edge~$(v,u)$ belongs to at most $d$ triangles. Applying this argument to
each of the three triangle edges gives $3d$.
Consequently, when sampling $pT$ triangles in expectation, an edge is deleted from at least $p  {T} / 3 d$
triangles (in expectation).  If $p T \in \Omega(\log n)$, the number of sampled
triangles is concentrated around the expectation (by Chernoff bounds). 

\subsubsection{Shortest Path Length} 

At most one edge is deleted from every triangle. Thus, the length of the
shortest $s$-$t$ path does not increase by more than  $2\times$ (in the
\textbf{worst case}), as we can always use the two edges remaining in the
triangle.

Moreover, we can show that the shortest $s$-$t$ path (previously of length
$\mathcal{P}$) has length at most $\mathcal{P}(1+p/3)$ \textbf{in expectation}.
As we consider each triangle for deletion at most once, the probability of
deleting an edge along the shortest path is at most $1/3$. Thus, we expect to
delete at most $p \mathcal{P}/3$ edges, increasing the length of the shortest
path by the same amount. 
This follows because along every path of length $\mathcal{P}$, the number of
deleted edges is stochastically dominated by a binomial random variable with
parameters $\mathcal{P}$ (number of trials) and $p/3$ (number of successes),
i.e., \#removed edges $\sim Bin \left(\mathcal{P}, p/3 \right)$. As each
deleted edge along the path increases the length of the shortest path by at
most one (again as we delete at most one edge from every triangle), the length
increases by at most the number of deleted edges along this path.
A similar reasoning gives the bounds for \emph{diameter}.

We can also obtain \textbf{high probability} concentration bounds by using
Chernoff bounds~\cite{boucheron2013concentration}, showing that a given shortest
path has length at most $\mathcal{P}(1 + p)$ w.h.p., if $\mathcal{P}$ is larger
than a constant times $\log n$.
Specifically, by a Chernoff bound, for any $0<\epsilon\le 1$, the probability
that more than $kp \epsilon /3$ edges are deleted along a length $k$ path
(i.e., a length~$k$ path has more than $kp (1+\epsilon)/3$ edges after sparsification) is
at most $e^{-kp \epsilon^2/9}$. Choosing $\epsilon=1$ and requiring $kp\geq 27
\log n$ gives us a probability of at most $n^{-3}$ that more than $\frac{2}{3} kp$
edges are deleted from a particular length~$k$ path. Bounding over all pairs of
vertices, we get that \emph{for all $s$-$t$ paths of length at least $(27 \log n)/
p$}, their lengths are preserved up to a factor $(1+ \frac{2}{3}p)$ with
probability at least $1-n^{-1}$.

In the above derivation, we rely on the fact that every edge is only
considered for deletion by at most one triangle, otherwise edges that are part
of more triangles are deleted with probability larger than $p/3$.

\subsubsection{Diameter} 
\ 
As a consequence of the bounds on the shortest path lengths, the diameter
increases by at most a factor two. Moreover, we can strengthen the result
similarly as for fixed paths and show that the diameter is at most $D + p D$
(w.h.p., as long as the diameter is larger than some constant times $\log n$).

We showed that for every $s$-$t$ pair the shortest path length increases by at
most a factor $(1+p)$ with probability at least $1-n^{-1}$.  Hence, the
diameter is at most $D + p D$ (w.h.p., as long as the diameter is larger than
some constant times $\log n$).

Note that the constants are not quite the same as where we considered just
individual paths: The problem lies in that very short paths could increase by a
factor two. The tighter bound on the diameter holds if the diameter is at least
twice the smallest size the $s$-$t$ paths are allowed to be. That is, the bound
holds if $D \geq 54 \log n$.

\subsubsection{Vertex Degree} 
\ 
A vertex of degree $d'$ is contained in at most $\lfloor d'/2 \rfloor$
edge-disjoint triangles. Hence, TR decreases its degree by at most $d'/2$.
As this bound holds for every vertex, it also holds for the
maximum degree and average degree.

\subsubsection{Chromatic Number} 
\ 
If the triangles where we delete edges from are \emph{vertex-disjoint}, we can
get bounds for the chromatic number that are nontrivial even when many
triangles are deleted.  We show that by deleting a set of vertex-disjoint edges
$E'$, the chromatic number decreases by at most a factor $2$. Say we are given
a coloring $C'$ of the compressed graph with $c'$ colors. Observe that the
graph $(V, E')$ consisting of the deleted edges $E'$ consists of connected
components that have $0$ or $1$ edge. Hence, this graph of deleted edges
trivially has a $2$-coloring $C''$. The idea is to construct a coloring from
$C'$ and $C''$ that respects both $C'$ and $C''$ using a kind of 'product'
construction. In the following, we think of the colors as integers $1, ..., c$.
We construct a coloring $C$ of the original graph from $C'$ as follows. Take
the coloring in compressed graph and apply it to the initial graph.  If the
color of a vertex $v$ in $C''$ is the first of the two colors, give $v$ the
same color as in $C'$. Otherwise, add $c'$ to the color of $v$ in $C'$. The
important thing to note is that two vertices receive the same color only if
they have the same color in both colorings $C'$ and $C''$. Now, we can see that
$C$ is a coloring of the original graph: Consider an arbitrary edge $e=(u, v)$.
If the edge $e$ is in the compressed graph, then $C'$ will assign $u$ and $v$
different colors. Otherwise, $e$ is in the set of deleted edges $E'$. As such,
$(u, v)$ receives distinct colors in $C''$. Therefore, the construction ensures
they receive distinct colors in $C$.  Clearly, the new coloring has at most
$2c'$ colors.
%

Similarly as for the chromatic number, if we delete a set of
\emph{vertex-disjoint} edges $E'$, the size of a maximum independent set
increases by at most a factor $2$. Consider an independent set of edges $I'$ of
the compressed graph of size $k'$. We construct an independent set $I$ of the
original graph of size at least $k'/2$, as follows. The graph $(V, E')$
consisting of the deleted edges is $2$ colorable. Choose the color for which
there are more vertices in $I'$ that have this color (pick either color if they
have the same number of vertices). We can see that this indeed gives an
independent set of the original graph and that it has size at least $k'/2$
(since one of the two colors must have the majority of the vertices in $I'$). 

\subsubsection{Maximum Cardinality Matching}\footnote{(11.2019) bound updated} 
\ 
In every triangle, a matching~\cite{besta2019substream} of the original graph
can contain at most one of its three edges. Since we delete at most one of the
three edges in a triangle uniformly at random, the probability that an edge in
a particular maximum matching of the original graph is deleted is at most
$1/3$. Hence, the expected number of edges that is deleted from the maximum
matching (originally of size $\hat M_C$) is at most $1/3 \hat M_C$.

\subsubsection{Coloring Number}
\ 
In a \emph{greedy coloring}, vertices are colored by visiting the vertices in
some predetermined ordering. The coloring number~\cite{erdHos1966chromatic}
gives the smallest number of colors obtained among all such vertex orderings by
a greedy coloring. This best ordering is closely related to the densest
subgraph, which is characterized by the
\emph{arboricity}~\cite{nash1961edge,DBLP:conf/isaac/ZhouN94}.

Let $m(S)$ be the number of edges in the subgraph of $G$ induced by the vertex
set $S$. The arboricity~\cite{nash1961edge} is given by $$\alpha =
\max_{\emptyset \subset S \subseteq V} \left\lceil \frac{ m(S) }{|S| - 1}
\right\rceil \enspace .$$ The arboricity relates to the coloring number $C_G$
by the inequalities $\alpha \leq C_G \leq 2\alpha
$~\cite{DBLP:conf/isaac/ZhouN94}.

\iftr 
Note that the coloring number is one plus the degeneracy of the graph.
\fi

Now, consider a set $S$ that obtains the maximum value. The expected number of
deleted edges from the subgraph induced by $S$ is at most $m(S)/3$. Hence, the
expected arboricity (and coloring number) of the compressed graph is at least
$\frac{2}{3}\alpha $, which is at least $\frac{1}{3}C_G$. 

\subsubsection{Others}
We observe that all connected components and the minimum spanning tree are preserved
(assuming that considered triangles are edge-disjoint and (in MST) the removed edge
has maximum weight in the triangle).

\iftr

\vspace{-0.5em}
\subsection{Edge Kernels: Spectral Sparsification}
\vspace{-0.25em}

We next consider spectral sparsification.

\iftr
The probability for an edge to be in the sparsified graph $p_{i,j}=\min(1, Y /
min(d_i, d_j))$. If $Y < 1$, vertices with the degree 1 can lose their adjacent
edges, that will harm the properties of the sparsifier
\cite{spielman2011spectral}. If $Y \ge d$ no edges will be removed. Therefore,
in the following derivations, we assume that the parameter $Y$ should be in
range $1 < Y < d$.

\subsubsection{Number of edges}

If all edges have an incident vertex with degree 1, then the probability for
the edge to appear in the new graph is $\min(1, Y) = 1$.  It is the upper
boundary.  Since vertices have degree at most $d$, the lower boundary is
$m(1-\min(1,Y/d)) = m \max(0, 1-Y/d) = m (1-Y/d)$.  Finally, $m(1-Y/d) \le
m^\star \le m$.

\fi

\macb{Connected Components} The number of connected components equals the multiplicity of the Laplacian eigenvalue $0$~\cite{spielman2011spectral,zumstein2005comparison}.
\iftr
This can be seen by two observations. First, the multiplicity of the Laplacian eigenvalue $0$ is one if and only if the graph is connected~\cite{spielman2011spectral,zumstein2005comparison}. Second, the spectrum of a graph with a set of connected components equals the union of the spectra of the connected components~\cite{zumstein2005comparison}.
\fi
As spectral sparsification preserves the eigenvalues up to a multiplicative factor (w.h.p.), the multiplicity of the eigenvalue $0$ does not change (w.h.p.).

\macb{Maximum Degree} The maximum degree is related to the largest Laplacian eigenvalue $\lambda_n$ by the inequality: $d\leq \lambda_n \leq 2d$~\cite{zumstein2005comparison}. Since the sparsified graphs approximates the eigenvalues well, we can show that the maximum degree of the sparsified graph is at least $d / 2(1+\epsilon) $. 

\iftr
 Let $d^{\star}$ be the maximum degree of the sparsified graph and $\lambda_n^{\star}$ bet its largest Laplacian eigenvalue. By applying the bound on the sparsified graph, we get $\lambda_n^{\star} \leq 2 d^{\star}$.

The sparsified graph approximates the Laplacian eigenvalues up to a factor $(1+\epsilon)$. Hence, we have that $$\lambda_n / (1+\epsilon) \leq \lambda_n^{\star} \leq (1+\epsilon)\lambda_n \enspace .$$ Combining this with the inequaility $d\leq \lambda_n \leq 2d$, we get$$d / (1+\epsilon) \leq \lambda_n^{\star} \leq (1+\epsilon) 2d$$ and finally (using $\lambda_n^{\star} \leq 2 d^{\star}$) we conclude $$d^{\star} \geq d / 2(1+\epsilon) \enspace . $$ 
\fi

\iftr
\macb{Chromatic Number} The bound on the chromatic number follows from the bound on the maximum degree.

\macb{Independent Set} The bound on the largest independent set follows from the bound on the chromatic number.
\fi

\fi

\subsection{Subgraph Kernels: Spanners}

Finally, we bound properties after using subgraph kernels.

\subsubsection{Number of Triangles}

A spanner consists of clusters that are connected with one another. Each
cluster is a tree, and every vertex has an edge to $O(n^{1/k})$ clusters (in
expectation)~\cite{miller2015improved}. Because the clusters are acyclic, a
triangle containing a vertex~$v$ has to contain one or two vertices that are in
a different cluster than $v$. There are $O(n^{2/k})$ possibilities to choose
two vertices in different clusters than $v$ (in expectation). Hence, summing
over all vertices, there are $O(n^{1+2/k})$ triangles \textbf{in expectation}.

\subsubsection{Minimum Coloring Number} 

Within each cluster, the edges form a tree. Any greedy coloring that colors
each of these trees bottom-up uses at most $O(n^{1/k}\log n)$ colors. We prove
this by bounding the number of edges to different clusters. 
%

We color vertices adjacent to intra- and inter-cluster edges separately. The
former is a forest and is therefore $2$-colorable.
For the latter, we bound its maximum degree.
Specifically, the probability that a vertex has an edge to more than $l$
clusters is at most $(1-n^{-1/k})^{l-1}$~\cite{miller2015improved}. Setting
$l=n^{1/k}2\log n + 1$ and using $1-x\leq e^x$, we get for the probability that
a fixed vertex has an edge to more than $l$ clusters:

$$
\left(1-n^{-1/k}\right)^{l-1} \leq e^{\frac{l-1}{n^{1/k}}} = n^{-2}.
$$

By a union bound over all vertices, the probability that a vertex has edges to
more than $l=O(n^{1/k}\log n)$ clusters is at most $n^{-1}$. 
Hence, the maximum number of edges from a vertex to vertices in other clusters
is $O(n^{1/k}\log n)$, meaning that it has a $O(n^{1/k}\log n)$-coloring. 
Hence, there is a greedy coloring which uses at most $O(n^{1/k}\log n)$ colors.
Combining the two colorings (intra-cluster trees and inter-cluster edges) gives
a coloring of the $O(k)$ spanner with $O(n^{1/k}\log n)$ colors.

\subsubsection{Maximum independent set} 
\ 
Since a $c$-colorable graph has an independent set of size at least $n/c$
(given by the vertices with the most frequent color), the $O(k)$ spanner has an
independent set of size at least $\Omega(n^{1-1/k}/\log n)$ (by our bound on
the chromatic number of the spanner).
%

\subsection{Subgraph Kernels: Lossy Summarization}

In \emph{Lossy Graph $\epsilon$-Summarization}, one bounds the size of the
symmetric difference between the adjacency lists of the compressed graph and
the original graph. That is, every vertex $v$ can have at most $\epsilon d_v$
neighbors whose connectivity to $v$ is incorrectly described by the compressed
graph (i.e., either there is an edge in the compressed graph even if there is
no such edge in the original graph, or vice versa). This means that edges can
be \emph{both added and removed} in lossy summaries. 

Immediately from the definition, we can see that the edge degrees of the
compressed graph are approximately the same as in the original graph. The
overall number of added (or removed) edges is at most $\Sigma_{v\in V}\
\epsilon d_v = 2\epsilon |E|$.  Then, the number of connected components can
change arbitrarily (when edges in a cut set are deleted) and distances can
change arbitrarily in the worst case (when an edge is added between distant
vertices or removal of edges disconnects the graph).  Other graph properties
may also change arbitrarily, possibly depending on $\epsilon$, see
Table~\ref{tab:theory-table}.

\iftr
\lukas{Note that particular graph summarization schemes might obtain better
bounds if they also restrict the type of edges that get added/removed, but for
the schemes I considered there are no claims to this end}
\fi

\begin{table*}[t]
\centering
\sf
\renewcommand{\arraystretch}{1}
\begin{tabular}{l}
\toprule
\makecell[l]{
\textbf{\ul{Friendships:}} 
%
 Friendster ({\textbf{s-frs}}, 64M, 2.1B), 
 Orkut ({\textbf{s-ork}}, 3.1M, 117M),
 LiveJournal ({\textbf{s-ljn}}, 5.3M,\\49M),
 Flickr ({\textbf{s-flc}}, 2.3M, 33M),
 Pokec ({\textbf{s-pok}}, 1.6M, 30M),
 Libimseti.cz ({\textbf{s-lib}}, 220k, 17M),\\
 Catster/Dogster ({\textbf{s-cds}}, 623k, 15M),
 Youtube ({\textbf{s-you}}, 3.2M, 9.3M),
 Flixster ({\textbf{s-flx}}, 2.5M, 7.9M),
%
%
%
%
}\\
\midrule
\makecell[l]{\textbf{\ul{Hyperlink graphs:}}
 Web Data Commons 2012 ({\textbf{h-wdc}}, 3.5B, 128B),
 EU domains (2015)\\({\textbf{h-deu}}, 1.07B, 91.7B),
 UK domains (2014) ({\textbf{h-duk}}, 787M, 47.6B),
 ClueWeb12 ({\textbf{h-clu}}, 978M, 42.5B),\\
 GSH domains (2015) ({\textbf{h-dgh}}, 988M, 33.8B),
 SK domains (2005) ({\textbf{h-dsk}}, 50M, 1.94B),\\
 IT domains (2004) ({\textbf{h-dit}}, 41M, 1.15B),
 Arabic domains (2005) ({\textbf{h-dar}}, 22M, 639M),\\
 Wikipedia/DBpedia (en) ({\textbf{h-wdb}}, 12M, 378M),
 Indochina domains (2004) ({\textbf{h-din}}, 7.4M, 194M),\\
 Wikipedia (en) ({\textbf{h-wen}}, 18M, 172M),
 Wikipedia (it) ({\textbf{h-wit}}, 1.8M, 91.5M),\\
 Hudong ({\textbf{h-hud}},  2.4M, 18.8M),
 Baidu ({\textbf{h-bai}}, 2.1M, 17.7M),
 DBpedia ({\textbf{h-dbp}}, 3.9M, 13.8M),
}\\
\midrule
\makecell[l]{\textbf{\ul{Communication:}}
 Twitter follows ({\textbf{m-twt}}, 52.5M, 1.96B),
 Stack Overflow\\interactions ({\textbf{m-stk}}, 2.6M, 63.4M),
 Wikipedia talk (en) ({\textbf{m-wta}}, 2.39M, 5M),
}\\
\midrule
\makecell[l]{\textbf{\ul{Collaborations:}}
 Actor collaboration ({\textbf{l-act}}, 2.1M, 228M),
 DBLP co-authorship ({\textbf{l-dbl}}, 1.82M,\\13.8M),
 Citation network (patents) ({\textbf{l-cit}}, 3.7M, 16.5M),
 Movie industry graph ({\textbf{l-acr}}, 500k, 1.5M)
}\\
\midrule
\makecell[l]{\textbf{\ul{Various:}}
 UK domains time-aware graph ({\textbf{v-euk}}, 133M, 5.5B),
 Webbase crawl\\({\textbf{v-wbb}}, 118M, 1.01B),
 Wikipedia evolution (de) ({\textbf{v-ewk}}, 2.1M, 43.2M),\\
 USA road network ({\textbf{v-usa}}, 23.9M, 58.3M),
 Internet topology (Skitter) ({\textbf{v-skt}}, 1.69M, 11M),
}\\
\bottomrule
\end{tabular}
\caption{\textmd{Considered
graphs with $n > 2$M or $m > 10$M from established datasets~\protect\cite{snapnets,
kunegis2013konect, 
demetrescu2009shortest, 
wdc, boldi2004webgraph}.
\textbf{Graph are sorted by $m$ in each category.}
For each graph, we show its ``(\textbf{symbol used later}, $n$, $m$)''.
\iftr
If there are multiple related graphs, for
example different snapshots of the .eu domain, we select the largest one. One
exception is the additional Italian Wikipedia snapshot selected due to its
interestingly high density.\fi
}}
%
\label{tab:graphs}
\end{table*}

\subsection{Discussion and Takeaways}

With simple \textbf{random uniform sampling} ($p$-sampling), the number of
connected components is not necessarily preserved. This means that the length
of a shortest path between any two vertices has unbounded expectation (i.e.,
the expectation does not exist). The advantage of $p$-sampling is
its simplicity. Moreover, it can be shown that if $p$ is large enough, the
compressed graph \emph{does} preserve the number of connected components with
large probability and the size of a minimum cut also obtains its expected value
in the sampled graph~\cite{karger00minimum}.
All other compression schemes considered preserve the number of connected
components, at least w.h.p..

\textbf{$O(k)$-Spanners} are designed to preserve the lengths of shortest paths (and
hence also the diameter), and this is what they do best. Spanners compress the
number of edges to close to linear in the number of vertices when a large
stretch~$k$ is allowed. However, for small stretch $k$ (e.g., $k=2$) the graph
can have many edges (up to $\min(m, n^{3/2})$). Interestingly, spanners also
allow for a coloring with relatively few colors and have a large independent
set. 

\textbf{Edge-Once Triangle $p$-$1$-Reduction} gives nontrivial bounds for
\emph{all} considered graph properties (except independent sets). Compressed
graphs are $2$-spanners and, w.h.p., $(\alpha=p, \beta=O(\log n))$-spanners.
Moreover, the compressed graph approximates the size of the {largest
matching} up to a factor $2/3$ and the {coloring number} up to a factor
$1/3$. If there are many triangles, the scheme can eliminate up to a third of
the {number of edges}. This is significant because $k$-spanners do not
guarantee compression for $k\leq 2$. 

\textbf{Spectral sparsification} preserves the value of {minimum cuts}
and {maximum flows}~\cite{karger00minimum,spielman2011spectral}.
Moreover, there is a relationship between the maximum degree of a graph and its
Laplacian eigenvalues, meaning that the {maximum degree} is preserved up
to a factor close to $2$. Thus, similarly to the original graph, the compressed
graph admits a {coloring} with $O(d)$ colors ($d$ is the maximum degree of the
original graph). Spectral sparsifiers always return a sparse graph, achieving a
{number of edges} that is close to linear in~$n$. 

\textbf{$\epsilon$-Summary} bounds the size of the {\emph{symmetric difference
between neighborhoods} in the compressed and original graph.} {Its bounds are
not competitive as this scheme can arbitrarily disconnect the graph and insert
new edges}, cf.~Table~\ref{tab:theory-table}.

\begin{figure*}[t]
%
  \centering
    \includegraphics[width=\textwidth]{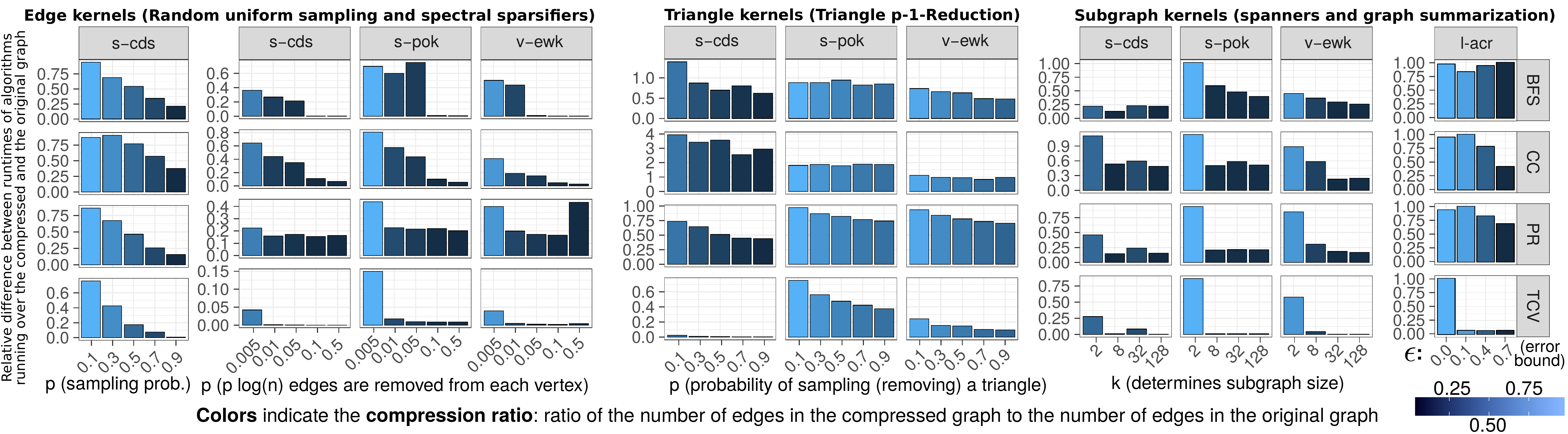}
%
%
\vspace{-1em}
%
\vspace{-1.5em}
\caption{\textmd{
  \textbf{Storage \& performance} tradeoffs of various lossy compression schemes implemented in Slim Graph (\textbf{when varying compression parameters}).
%
}}
%
%
%
%
\label{fig:storage-performance}
\end{figure*}

\section{Evaluation}
\label{sec:evaluation}

Lossy graph compression enables tradeoffs in three key aspects of graph
processing: \emph{performance}, \emph{storage}, and \emph{accuracy}.
We now illustrate several of
these tradeoffs.
\iftr
Numerous introduced schemes and extensibility of Slim Graph enable arbitrary
tradeoffs in the space of the above aspects.  We now illustrate several of
these tradeoffs.
\fi
\emph{Our goal is \textbf{not} to advocate a single compression
scheme, but to (1) confirm pros and cons of different schemes, provided in~\cref{sec:theory}, and
(2) illustrate that \textbf{Slim Graph 
enables analysis of the associated tradeoffs}.}

\macb{Algorithms, Schemes, Graphs}
We consider algorithms and compression schemes
from~\cref{sec:back} and Table~\ref{tab:theory-schemes}, and all associated parameters.
We also consider all large graphs from SNAP~\protect\cite{snapnets},
KONECT~\protect\cite{kunegis2013konect},
DIMACS~\protect\cite{demetrescu2009shortest}, Web Data
Commons~\protect\cite{wdc}, and WebGraph
datasets~\protect\cite{boldi2004webgraph};
see Table~\ref{tab:graphs} for details.
\emph{This creates a \textbf{very large evaluation space} and we only summarize
selected findings; full data is in the extended report.}


\iftr
\macb{Implementation Details}
We use graph algorithms implemented in the GAP Benchmark
Suite~\cite{beamer2015gap} and GraphChi~\cite{kyrola2012graphchi} for graphs
that fit and do not fit in memory. 
\fi

\iftr
\begin{figure*}
  \centering
  \begin{subfigure}[t]{0.21 \textwidth}
    \centering
    \includegraphics[width=\textwidth]{plots/exchange/exchange_lossy_4.eps}
    \caption{4 compute nodes.}
    \label{fig:x}
  \end{subfigure}
  \begin{subfigure}[t]{0.21 \textwidth}
    \centering
    \includegraphics[width=\textwidth]{plots/exchange/exchange_lossy_32.eps}
    \caption{32 compute nodes.}
    \label{fig:x}
  \end{subfigure}
  \begin{subfigure}[t]{0.21 \textwidth}
    \centering
    \includegraphics[width=\textwidth]{plots/exchange/exchange_lossy_256.eps}
    \caption{256 compute nodes.}
    \label{fig:x}
  \end{subfigure}
  \begin{subfigure}[t]{0.21 \textwidth}
    \centering
    \includegraphics[width=\textwidth]{plots/exchange/exchange_lossy_1024.eps}
    \caption{1024 compute nodes.}
    \label{fig:x}
  \end{subfigure}
%
\caption{The analysis on the amount of communicated data in a distributed-memory BFS for the lossy Slim Graph.}
\label{fig:lossy-dist}
\end{figure*}
\fi

\macb{Evaluation Methodology}
For algorithmic execution we use the arithmetic mean for data summaries. We treat the first 1\% of any
performance data as warmup and we exclude it from the results. We gather enough
data to compute the mean and 95\% non-parametric confidence intervals.

\macb{Machines}
We use
{CSCS Piz Daint}, a Cray with various XC* nodes. Each XC50 compute node
contains a 12-core HT-enabled Intel Xeon E5-2690 CPU with 64 GiB RAM.  Each
XC40 node contains two 18-core HT-enabled Intel Xeons E5-2695 CPUs with 64 GiB
RAM.
%
%
The interconnection is Cray's Aries and it implements the
Dragonfly topology~\cite{dally08}. The batch system is slurm 14.03.7.
This machine represents massively parallel HPC machines.
We also use high-end servers, most importantly a system with
Intel Xeon Gold 6140 CPU @ 2.30GHz,
768GB DDR4-2666,
18 cores, and
24.75MB L3.
\subsection{Storage and Performance}
\label{sec:eval-storage-perf}

We start with storage and performance tradeoffs.
Figure~\ref{fig:storage-performance} shows the impact of different
compression parameters on $m$ and performance (we use smaller graphs to
analyze in detail a large body of parameters).  Plotted graphs are selected to
cover different edge sparsity and number of triangles per vertex ($T/n$ is 1052
(s-cds), 20 (s-pok), and 80 (v-ewk)).
In most cases, \emph{spanners and $p$-1-TR ensure the largest and smallest \textbf{storage
reductions}}, respectively. This is because subgraphs in spanners become
spanning trees while $p$-1-TR removes only as many edges as the count of
triangles.  Uniform and spectral sampling offer a middle ground --- depending
on $p$, they can offer arbitrarily small or large reductions of $m$.
Moreover, \emph{respective storage reductions entail similar \textbf{performance effects}
(fewer edges indicates faster algorithmic execution).}
Still, there are some effects specific to each scheme. Spanners offer mild performance
improvements for small $k$ that increase by a large factor after a certain threshold of $k$
is reached.
Other schemes steadily accelerate all algorithms with growing $p$.
%
%
As expected, random uniform sampling ensures linear (with respect to~$p$) advantages in both
storage and performance.

\begin{figure}[h]
  \centering
    \includegraphics[width=0.49\textwidth]{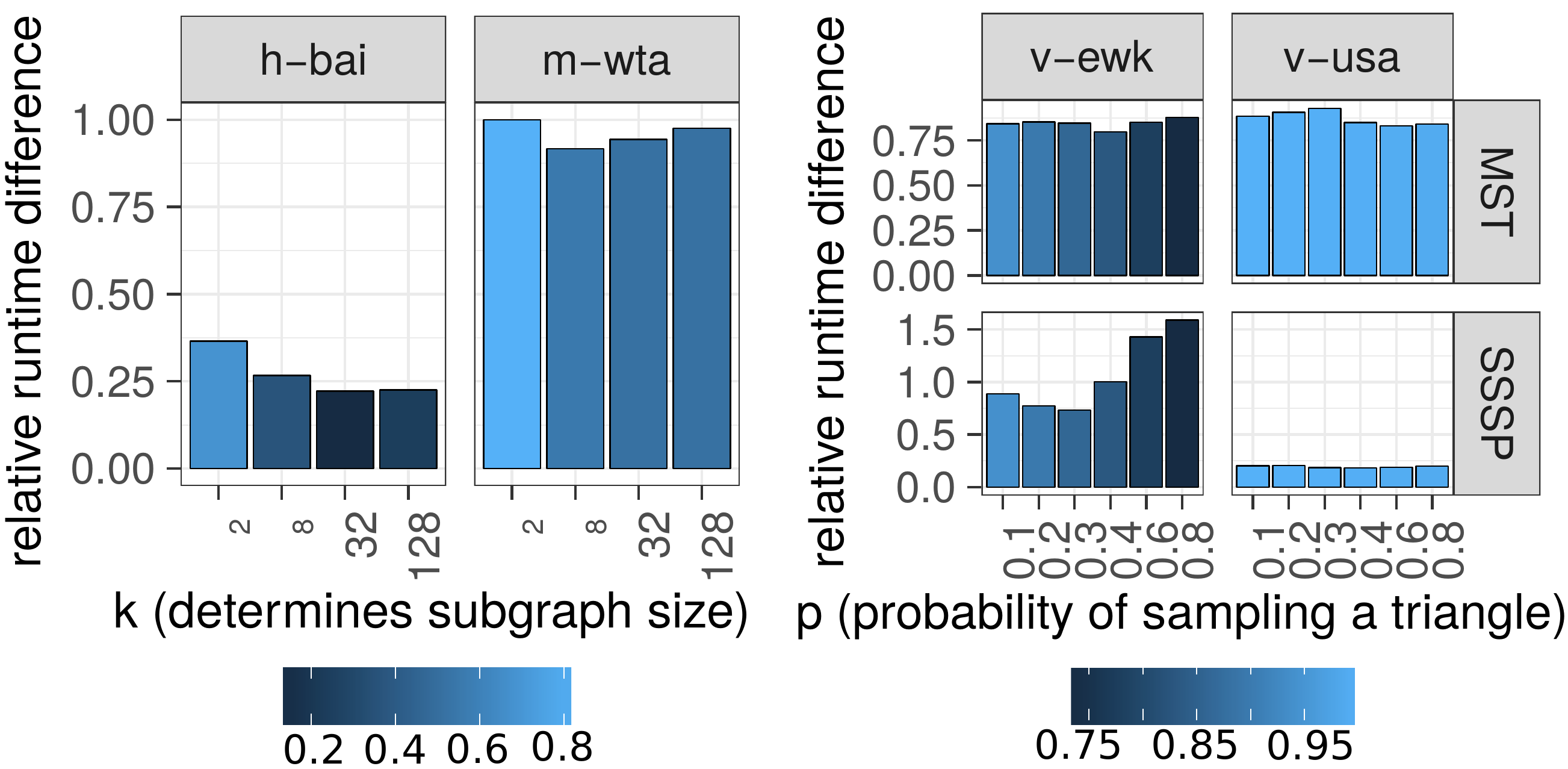}
%
%
%
\vspace{-2em}
\caption{\textmd{
 %
  \textbf{Storage and performance analysis:} performance of MCM (left) and SSSP / MST compressed with $p$-1-TR (right).
Colors indicate the compression ratio: ratio of the number of edges in the compressed graph to the number of edges in the original graph. 
%
}}
%
%
%
%
\label{fig:weighted}
\end{figure}

We also test TR on weighted graphs (resulted excluded due to space constraints), see Figure~\ref{fig:weighted}.
For very sparse graphs, such as the US road network, compression ratio and thus speedups (for both MST and SSSP) from TR is very low.
MST's performance is in general not influenced much because it depends mostly on $n$.
In other graphs, such as v-ewk, SSSP
speedups follow performance patterns for BFS.
For some graphs and roots, very high $p$ that significantly enlarges diameter
(and iteration count)
may cause slowdowns.
Changing $\Delta$ can help but needs manual tuning.
Consequently, we conclude that
\emph{lossy compression may also degrade performance
if a selected scheme is unsuitable for targeted algorithms.}


\iftr
 until a certain threshold ($p=3$ for v-ewk).
Higher $p$ causes slowdowns because the SSSP iteration count grows when many edges
are deleted; this is particularly the case for SSSP as each step in the
underlying $\Delta$-Stepping~\cite{meyer2003delta_} implementation is associated with edge relaxation using buckets,
which may be expensive.
\emph{Thus, lossy compression may also degrade performance if
a graph is compressed with a scheme unsuitable for the following algorithms.}
\fi

\begin{figure}[h]
%
  \centering
    \includegraphics[width=0.28\textwidth]{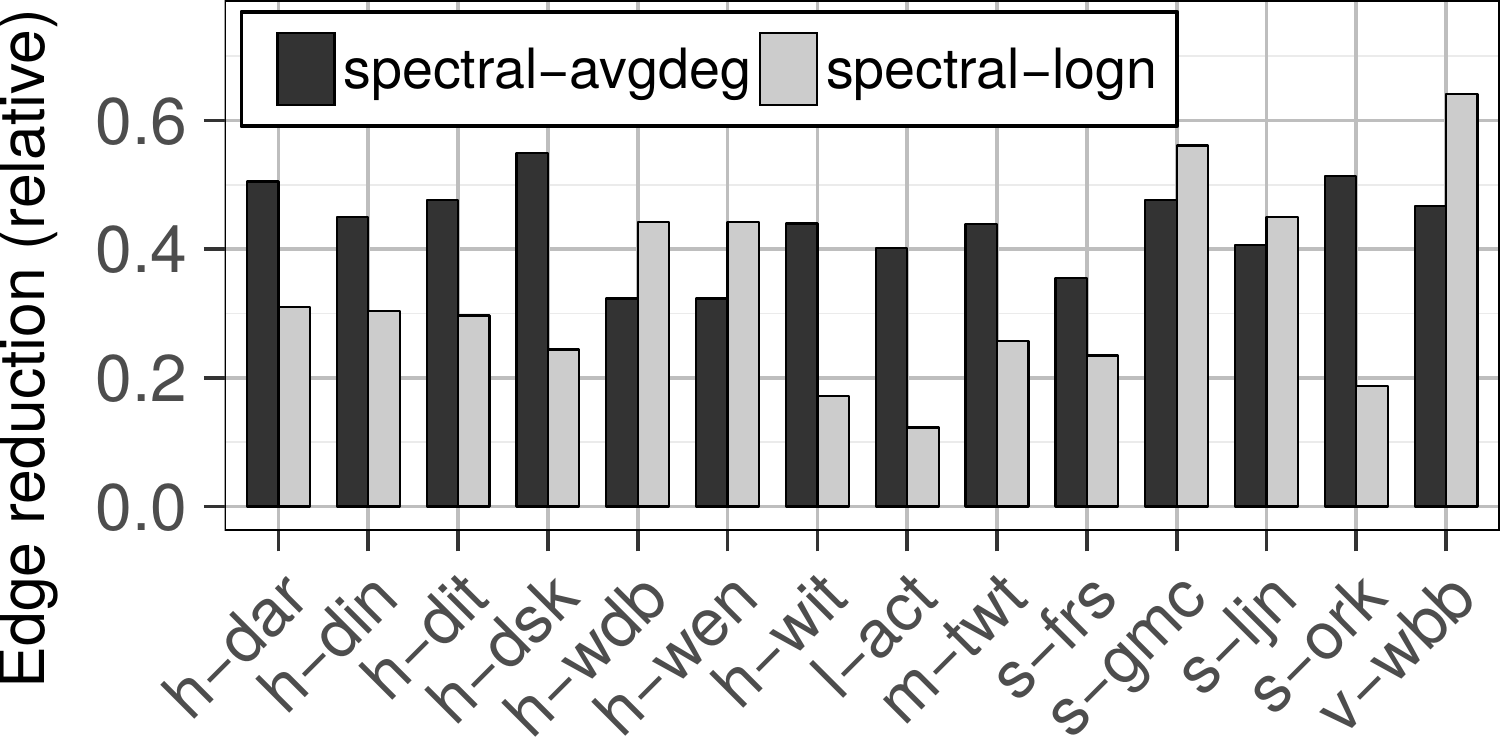}
    \includegraphics[width=0.185\textwidth]{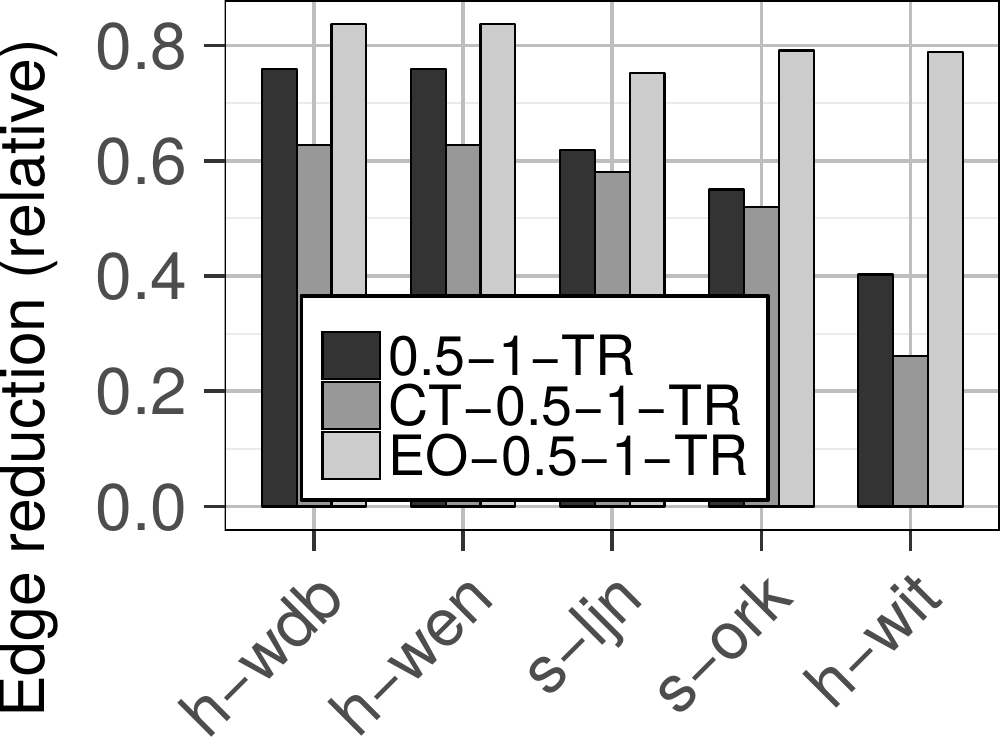}
%
%
%
\vspace{-1.25em}
\caption{\textmd{
  \textbf{Compression ratio analysis}: different variants of spectral sparsification (left) and triangle reduction (right), for a fixed $p=0.5$.
  Extending results from Figure~\ref{fig:storage-performance} (panels ``spectral sparsification'' and ``TR'', argument $p=0.5$)
  to (1) graphs of different sizes, sparsities, classes, degree distributions, and (2) multiple compression variants}.
%
}
%
%
%
%
\label{fig:size-variants}
\end{figure}

We also analyze variants of proposed Slim Graph compression kernels. Figure~\ref{fig:size-variants}
shows size reductions in graphs compressed with spectral sparsification variants, in which
the number of remaining edges is proportional to the average degree or $\log(n)$.
We also analyze variants of TR; ``CT'' is an additional variant of ``EO'' in which
we not only consider an edge for removal \emph{at most once}, but also
\emph{we remove edges starting from ones that belong to the fewest triangles}.
Spectral variants result in different size reductions, depending on graphs.
\emph{Contrarily, the ``CT'' and ``EO'' TR variants consistently deliver smaller
$m$ than a simple $p$-1-TR (for a fixed $p=0.5$).}

\subsection{Accuracy}

We use Slim Graph metrics to analyze the accuracy of graph algorithms after
compressing graphs.
First, we show that the Kullback-Leibler divergence can assess information loss
due to compression, see Table~\ref{tab:pr-KL}.  \emph{In all the cases, the
higher compression ratio is (lower $m$), the higher KL divergence becomes.}

\begin{table}[h]
\centering
\sf
\footnotesize
%
\setlength{\tabcolsep}{2pt}
\renewcommand{\arraystretch}{1}
\begin{tabular}{ccccccccc@{}}
\toprule
\textbf{Graph} & \makecell[c]{EO\\0.8-1-TR} & \makecell[c]{EO\\1.0-1-TR} & \makecell[c]{Uniform\\($p=0.2$)} & \makecell[c]{Uniform\\($p=0.5$)} & \makecell[c]{Spanner\\($k=2$)} & \makecell[c]{Spanner\\($k=16$)} & \makecell[c]{Spanner\\($k=128$)}\\
\midrule
s-you &  0.0121 & 0.0167 & 0.1932 & 0.6019 & 0.0054 & 0.2808 & 0.2993\\
h-hud &  0.0187 & 0.0271 & 0.0477 & 0.1633 & 0.0340 & 0.2794 & 0.3247 \\
l-dbl &  0.0459 & 0.0674 &  0.0749 & 0.2929 & 0.0080 & 0.1980 & 0.2005 \\
v-skt &  0.0410 & 0.0643  & 0.0674 & 0.2695 & 0.0311 & 0.1101 & 0.2950 \\
v-usa &  0.0089 & 0.0100  & 0.1392 & 0.5945 & 0.0000 & 0.0074 & 0.0181 \\

\bottomrule
\end{tabular}
\caption{\textmd{\textbf{Kullback-Leibler divergences} between
PageRank probability distributions on the
original and compressed graphs, respectively.}}
%
\label{tab:pr-KL}
\end{table}

\iftr
\maciej{fix}
Another proposed metric, $\mathcal{A}_{REN}$, is the number of pairs of neighboring vertices that swapped their
\fi

Another proposed metric is the number of pairs of neighboring vertices that swapped their
order (with respect to a certain property) after compression.
We test this metric for BC and TC per vertex.
Note that this metric should be used
\emph{when the compared schemes remove the same number
of edges (possibly in expectation)}. Otherwise, numbers of reordered vertices
may differ simply because one of compared graphs has fewer vertices left.
With this metric, we discover that \emph{spectral sparsification
preserves TC per vertex better than other methods}.


\iftr
\maciej{fix}
With $\mathcal{A}_{REN}$, we discover that \emph{spectral sparsification
\fi

We also discover that used $O(k)$-spanners preserve the accuracy of the BFS
traversal trees surprisingly well.  For example, for the s-pok graph,
respectively, removing 21\% ($k=2$), 73\% ($k=8$), 89\% ($k=32$), and 95\%
($k=128$) of edges preserves 96\%, 75\%, 57\%, and 27\% of the critical edges
that constitute the BFS tree.
\emph{The accuracy is maintained when different root vertices are picked and
different graphs are selected}.

We also investigate how triangle count ($T$) is reduced with lossy compression.
Intuitively, TR should significantly impact $T$.
While this is true, we also illustrate that \emph{almost all schemes, especially spanners,
eliminate a large fraction of triangles}, see Table~\ref{tab:tcv}.
This is because spanners, especially for large $k$, remove most of cycles
while turning subgraphs into spanning trees.

\begin{table}[h]
\centering
\sf
\ssmall
\setlength{\tabcolsep}{2pt}

\begin{tabular}{ccccccccccccc@{}}
\toprule
\textbf{Graph} & \begin{turn}{90}\makecell[c]{Original}\end{turn} & \begin{turn}{90}\makecell[c]{0.2-1-TR}\end{turn} & \begin{turn}{90}\makecell[c]{0.9-1-TR}\end{turn} & \begin{turn}{90}\makecell[c]{Uniform\\($p=0.8$)}\end{turn} & \begin{turn}{90}\makecell[c]{Uniform\\($p=0.5$)}\end{turn} & \begin{turn}{90}\makecell[c]{Uniform\\($p=0.2$)}\end{turn} & \begin{turn}{90}\makecell[c]{Spanner\\($k=2$)}\end{turn} & \begin{turn}{90}\makecell[c]{Spanner\\($k=16$)}\end{turn} & \begin{turn}{90}\makecell[c]{Spanner\\($k=128$)}\end{turn} & \begin{turn}{90}\makecell[c]{Spectral\\($p=0.5$)}\end{turn}&  \begin{turn}{90}\makecell[c]{Spectral\\($p=0.05$)}\end{turn} & \begin{turn}{90}\makecell[c]{Spectral\\($p=0.005$)}\end{turn}\\
\midrule
%
%
s-you &  11.38 & 1.544 & 0.037 & 0.091 & 1.416 & 5.825 & 7.626 & 0.071 & 0.000 & 0 & 0.007 & 0.426\\
s-flx &  9.389 & 0.645 & 0.017 & 0.075 & 1.173 & 4.802 & 6.933 & 0.000 & 0.070 & 0 & 0.001 & 0.219\\
s-flc &  1091 & 6.845 & 0.164 & 8.765 & 136.6 & 557.9 & 250.7 & 1.327 & 0.001 & 0 & 0.016 & 1.517\\
s-cds &  3157 & 18.56 & 0.561 & 25.24 & 394.8 & 1615 & 844.5 & 45.392 & 0.001 & 0 & 0.015 & 4.821\\
s-lib &  938.3 & 31.51 & 0.902 & 7.569 & 116.9 & 480.2 & 82.59 & 167.0 & 5.708 & 0 & 0.000 & 0.042\\
s-pok &  59.82 & 10.25 & 0.280 & 0.480 & 7.494 & 30.58 & 41.27 & 0.362 & 0.000 & 0 & 0.005 & 1.962\\
h-dbp &  6.299 & 1.158 & 0.072 & 0.051 & 0.822 & 3.218 & 2.295 & 0.440 & 0.002 & 0 & 0.020 & 1.981\\
h-hud &  14.71 & 1.832 & 0.083 & 0.117 & 1.839 & 7.538 & 7.373 & 0.001 & 0.000 & 0 & 0.005 & 2.495\\
l-cit &  5.973 & 1.994 & 0.091 & 0.048 & 0.747 & 3.059 & 5.128 & 0.240 & 0.000 & 0 & 0.007 & 1.931\\
l-dbl &  45.57 & 6.144 & 0.257 & 0.365 & 5.671 & 23.33 & 22.64 & 0.033 & 0.004 & 0 & 0.066 & 8.572\\
v-ewk &  235.2 & 14.13 & 0.422 & 1.886 & 29.33 & 120.3 & 110.0 & 0.034 & 0.000 & 0 & 0.008 & 2.436\\
v-skt &  50.88 & 2.642 & 0.099 & 0.395 & 6.455 & 26.01 & 22.24 & 5.777 & 0.502 & 0 & 0.016 & 2.376\\
%
%
%
\bottomrule
\end{tabular}
\caption{(Accuracy) \textmd{Analysis of the average number of triangles per vertex.}}\label{tab:tcv}
%
\end{table}

Further tradeoffs between accuracy and size reductions are
related to other graph properties.
For example, the MM size is least affected by TR. Similarly, the
MST is preserved best by TR (assuming a variant that always removes
the maximum weight edge in a triangle), followed by spanners.
In SSSP, spanners best preserve lengths of shortest paths,
followed by TR.
Finally, 
spanners and the ``EO'' variant of TR maintain the
number of CC. Contrarily, random uniform sampling
and spectral sparsification disconnect graphs. Graph summarization
acts similarly to random uniform sampling (also with respect to other properties), because it can also
arbitrarily remove edges. However, for a fixed $p$,
the latter generates significantly fewer (by $>$10$\times$) components than the former;
this is because \emph{used spectral sparsification schemes were designed to minimize
graph disconnectedness}.



\begin{figure*}[hbtp]
%
  \centering
    \includegraphics[width=0.9\textwidth]{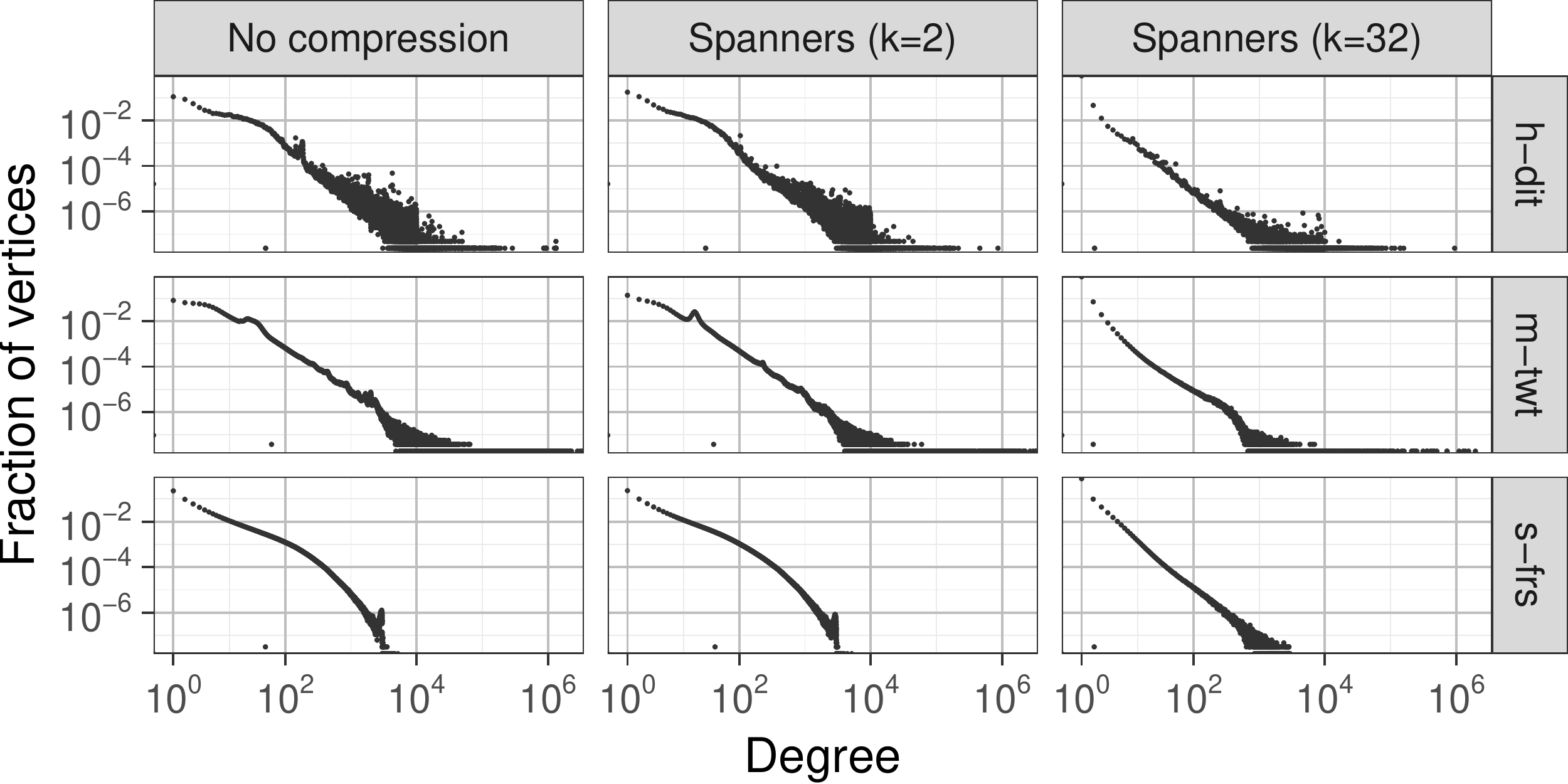}
%
%
%
%
%
\caption{\textmd{
  \textbf{Accuracy analysis (\textbf{varying $k$})}: impact of spanners on the degree distribution of popular graph
  datasets, Twitter communication (m-twt), Friendster social network (s-frs), and~.it domains (h-dit).
  Extending results from Figure~\ref{fig:storage-performance} (panel ``spanners'', arguments $k \in \{2,32\}$)
  to degree distribution}.
%
}
%
%
%
%
\label{fig:deg-dists}
\end{figure*}

In Slim Graph, we also analyze the impact of compression kernels on degree
distributions. As degree distributions determine many structural and
performance properties of a graph, \emph{such analysis is a visual
method of assessing the impact of compression on the graph structure}. This method is
also \emph{applicable to graphs with different vertex counts}. We
illustrate the impact from spanners on three popular graphs often used in
graph processing works (Twitter, Friendster,~.it domains) in
Figure~\ref{fig:deg-dists}.
Interestingly, spanners ``strengthen the power law'': the higher $k$ is,
the closer to a straight line the plot is.
\emph{One could use such observations to accelerate graph processing frameworks
that process compressed graphs, by navigating the design of data distribution
schemes, load balancing methods, and others.}

\begin{figure*}[hbtp]
%
  \centering
    \includegraphics[width=0.9\textwidth]{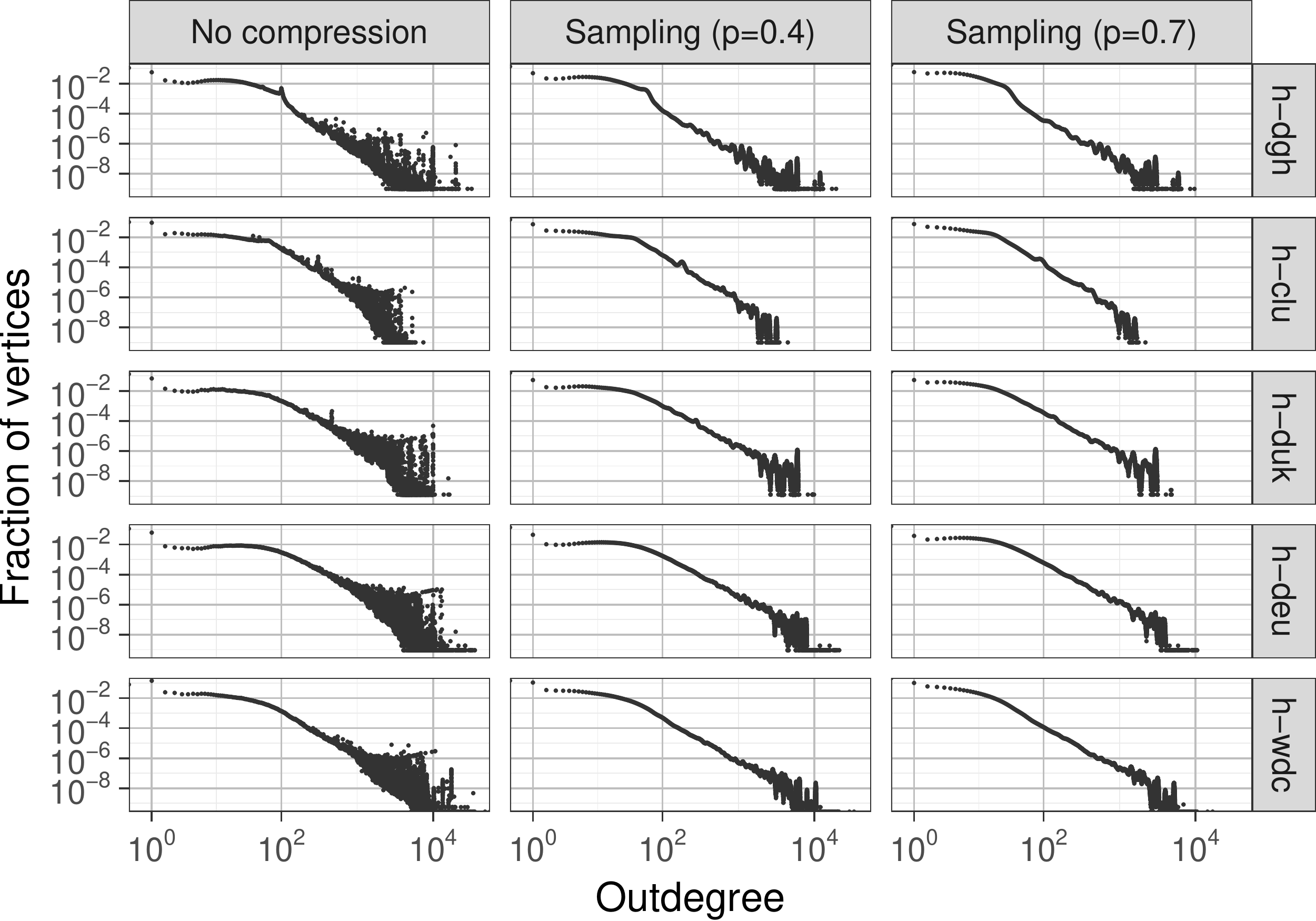}
%
%
%
%
%
\caption{\textmd{
  \textbf{(Accuracy)} Impact of random uniform sampling on the degree distribution of
large graphs (the largest, h-wdc, has $\approx$128B edges).
\#Compute nodes used for compression: 100 (h-wdc), 50 (h-deu), 20 (h-duk), 13 (h-clu), and 10 (h-dgh).
%
}}
%
%
%
%
\label{fig:dist-degree-dists}
\end{figure*}


\subsection{Distributed Compression of Large Graphs}
\label{sec:eval-dist}

To the best of our knowledge, \emph{we present the first results from distributed
lossy graph compression}.
In a preliminary analysis, we compressed the \emph{five largest publicly available
graphs} using edge kernels (random uniform sampling) and we analyze 
their degree distributions in Figure~\ref{fig:dist-degree-dists}.
%
%
Random uniform sampling ``removes the clutter'': scattered points that
correspond to specific fractions of vertices with
different degrees. 
\emph{This suggests that random uniform sampling could be used
as preprocessing for more efficient investigation into
graph power law properties.}


\subsection{Other Analyses}
\label{sec:eval-others}


We also compared Slim Graph kernels against 
low-rank approximation (of adjacency or Laplacian graph matrix).
It entails significant storage overheads
(cf.~Table~\ref{tab:theory-schemes}) and consistently very high
error rates. 
%
%
We also \textbf{timed the compression routines}. The compression time
is {not} a bottleneck and it follows asymptotic complexity 
($O(m)$ for uniform sampling, spectral sparsification, and spanners, $O(Im)$ for summarization, and
$O(m^{3/2})$ for TR). In all cases, sampling
is the fastest; spectral sparsification is negligibly slower as each kernel
must access degrees of attached vertices. Spanners are $>$20\% slower
due to overheads from low-diameter decomposition (larger constant factors in $O(m)$).
TR is slower than spanners by $>$50\% ($O(m^{3/2}$ vs.~$O(m)$).
Summarization is $>$200\% slower than TR due to large
constant factors and a complex design.
%

\subsection{How To Select Compression Schemes?}


We \textbf{summarize our analyses} by providing guidelines on selecting a compression scheme
for a specific algorithm.
Overall, \textbf{empirical analyses follow our theoretical predictions}.
Thus, as \textbf{the first step}, we recommend
to consult
Table~\ref{tab:theory-table} and select a compression scheme that ensures best
accuracy.
\textbf{Second}, one should verify whether a selected
method is feasible, given the input graph size \emph{and} graph type, e.g., whether a scheme
supports weighted or directed graphs. Here, we offer Table~\ref{tab:theory-schemes}
for overview and Section~\ref{sec:eval-others} with remarks on
  empirical performance.
\textbf{Third}, to select concrete parameter values, one should consult Figure~\ref{fig:storage-performance},
key insights from~\cref{sec:eval-storage-perf}--\cref{sec:eval-dist},
and -- possibly -- the report with more data.

\section{Related Work}
\label{sec:rw}

We now briefly
discuss related works. 

\macb{Lossy graph compression }
is outlined in~\cref{sec:back}, \cref{sec:sg-ot}, and in
Table~\ref{tab:theory-schemes}. \emph{We analyze its feasibility
for practical usage and we express and implement representative schemes
  as Slim Graph compression kernels}, covering spanners~\cite{peleg1989graph},
  spectral sparsifiers~\cite{spielman2011spectral}, graph
  summarization~\cite{shin2019sweg}, and others~\cite{maserrat2012community}.
  Our TR schemes generalize past work that removes two edges from
  triangles in weighted graphs to preserve exact shortest
  paths~\cite{kalavri2016shortest}. \emph{Most remaining schemes could
  be implemented as Slim Graph kernels.}

Second, \macb{lossless graph compression} is summarized in a recent
survey~\cite{besta2018survey}; it is outside the Slim Graph scope.

Third, many \macb{approximation graph algorithms} (i.e., graph algorithms with
provable approximation ratios) have been develop to alleviate the hardness of
various NP-Complete, NP-Hard, and related graph problems. Examples include
balanced cuts~\cite{even1999fast}, graph coloring~\cite{halldorsson1993still,
de2004approximate}, vertex covers~\cite{khot2008vertex}, solving the Traveling
Salesman Problem~\cite{christofides1976worst}, and many
more~\cite{de2004approximate, wang1995algorithms}. Contrarily to Slim Graph,
\emph{these works are usually sophisticated theoretical designs that are hard
to use in practice}; many of them focus on techniques for proving approximation
hardness. Moreover, \emph{they do not consider compressing input graphs and
thus they are orthogonal to Slim Graph and could be combined with our work for
more performance}.

More recently, there have been several attempts at \macb{approximating graph
computations} without explicitly considering the compression of the underlying
graphs. As opposed to the traditional ``Approximation Algorithms'', these works
do not specifically tackle problems that are ``hard'' in the formal sense.  The
  vast majority of these works are dedicated to a single algorithm or problem
  that poses computational difficulties even if it is in the P class (because
  of, for example, quadratic complexity).  Examples are betweenness
  centrality~\cite{riondato2016fast, borassi2016kadabra, riondato2018abra,
  geisberger2008better, bader2007approximating, chehreghani2018efficient},
  minimum spanning tree weight~\cite{chazelle2005approximating},
  reachability~\cite{dumbrava2018approximate}, motif
  counting~\cite{iyer2018asap, slota2014complex}, graph
  diameter~\cite{chechik2014better, roditty2013fast}, and
  others~\cite{roditty2013fast, boldi2011hyperanf, echbarthi2017lasas}.  The
  vast majority of these algorithms uses some form of sampling.  For example,
  works on betweenness centrality usually sample shortest paths.
Moreover, preliminary attempts were made at general approximate graph
processing~\cite{shang2014auto, iyer2018bridging, singh2018scalable}.

%
%
%

Finally, there exist various works at \macb{general approximate
processing}~\cite{han2013approximate, mittal2016survey}.  They relax the need
for full precision at the level of arithmetic blocks, processing units, pieces
  of code, pertinent error and quality measures, algorithms, programming
  models, and many others.

\iftr
\noindent
\ding{184} \macb{Cut Sparsifiers}
In cut sparsification~\cite{benczur1996approximating}, one approximates an 
input graph~$G$ with a graph~$H$ such that, for every subset of vertices $U
\subset V$, the weight of the edges leaving $U$ is approximately the same in
$G$ as in the sparsifier $H$.  Cut sparsifiers are a special case of spectral
sparsifiers: a given $G$ and its cut sparsifier $H$ must satisfy the same set
if inequalities as in spectral sparsification, but only for $x \in \{0,1\}^n$.
\textbf{We do not use cut sparsifiers as they are NP-Hard to derive}.

\noindent
\ding{185} \macb{Low-Rank Approximation of Graphs}
In low-rank approximation of graphs, 
one does \emph{not} explicitly remove edges. Instead,
\fi

\iftr
one first clusters the input graph~$G$ and
then approximates each cluster separately using existing
methods, e.g. the singular value decomposition, or stochastic
algorithms. The cluster-wise approximations are then
extended to approximate the entire graph. This approach
has several benefits: (1) important community structure of
the graph is preserved due to the clustering; (2) highly accurate
low rank approximations are achieved; (3) the procedure
is efficient both in terms of computational speed and memory
usage; (4) better performance in problems from various
applications compared to standard low rank approximation.
Further, we generalize stochastic algorithms to the clustered
low rank approximation framework and present theoretical
bounds for the approximation error. 
\fi

\section{Conclusion}

We introduce Slim Graph: the first framework and programming model for lossy graph compression. 
The core element of this model are \emph{compression
kernels}: small code snippets that modify a local part of the graph, for
example a single edge or a triangle.  Compression kernels can express and
implement multiple methods for lossy graph compression, for example spectral
sparsifiers and spanners. To ensure that Slim Graph is versatile, we
exhaustively analyzed a large body of works in graph compression theory.
Users of Slim Graph could further extend it towards novel compression methods.

Slim Graph introduces metrics for assessing the
quality of lossy graph compression. Our metrics target different classes of
graph properties, e.g., vectors of numbers associated with each vertex,
or probability distributions. For the latter, we propose to use statistical
divergences, like the Kullback-Leibler divergence, to evaluate
information loss caused by compression. Slim Graph could be
extended with other metrics.

In theoretical analysis, we show how different compression methods impact
different graph properties. We illustrate or derive more than 50 bounds. For
example, we constructively show that a graph compressed with Triangle Reduction (TR)
has a maximum cardinality matching (MCM) of size at least half of the size of
MCM in the uncompressed graph. TR is a novel class of
compression methods, introduced in Slim Graph, that generalizes past work and is \emph{flexible}: one can easily
tune it to preserve accurately various graph properties.

We use Slim Graph to evaluate different schemes in terms of
(1) reductions in graph sizes, (2) performance of algorithms running
over compressed graphs, and (3) accuracy in preserving  graph properties.
We also conduct the first distributed lossy compression
of the largest publicly available graphs.
We predict that \emph{Slim Graph may become a platform for
designing and analyzing today's and future lossy graph compression
methods, facilitating approximate graph processing, storage,
and analytics}.

{\small \section*{Acknowledgments}
%
%
We thank Mark Klein, Hussein Harake, Colin McMurtrie, and the whole CSCS team
granting access to the Ault and Daint machines, and for their excellent
technical support.  We thank David Schmidig for help with analyzing low-rank
approximation, and Timo Schneider for his immense help with computing
infrastructure at SPCL.  This project has received funding from the European
Research Council (ERC) under the European Union's Horizon2020 programme (grant
agreement DAPP, No.678880) and Google (European Doctoral Fellowship).  }

\balance

\bibliographystyle{ACM-Reference-Format}
\bibliography{references}

\end{document}